\documentclass[aps,prd,reprint,groupedaddress,amsmath,amssymb]{revtex4-2}

\usepackage{graphicx}
\usepackage{bm}
\usepackage{amssymb}
\usepackage{amsmath}
\usepackage{color}
\usepackage{mathtools}
\usepackage{cases}
\usepackage{booktabs}
\usepackage{mathrsfs}
\usepackage{float}
\usepackage{array}
\usepackage{dcolumn}
\usepackage{bm}
\usepackage{braket}

\usepackage[
colorlinks=true,
filecolor=black,
anchorcolor=blue,
linkcolor=blue,
citecolor=cyan, 
urlcolor=cyan,
linktocpage=true,
plainpages=false,
breaklinks=true,
pdfstartview=FitH
]{hyperref}

\newcommand{\mmn}{M_{\mu\nu}}
\newcommand{\mn}{\mu\nu}
\newcommand{\mij}{M_{ij}}
\newcommand{\dm}{\mathrm{d}}

\newcommand{\calG}{{g}}
\newcommand{\mpl}{M_{\rm Pl}}
\newcommand{\tg}{\tilde{g}}

\newcommand{\rhodm}{\rho_{\mathrm{DM}}}

\newcommand{\beam}{F}

\newcommand{\detector}[1]{{\rm D}_{#1}^{ij}(\vec{v})}
\newcommand{\nsr}{\hat{n}_{rs}}
\newcommand{\transfer}{\mathcal{T}}
\newcommand{\doppler}[1]{(1+\vec{v}\cdot \hat{n}_{#1})}
\newcommand{\doppleranti}[1]{(1-\vec{v}\cdot \hat{n}_{#1})}
\newcommand{\response}{\mathcal{R}}
\newcommand{\trls}[1]{(\vec{v}\cdot \hat{n}_{#1})}

\newcommand{\phim}{\phi_m}

\begin{document}
\title{Probing Spin-2 Ultralight Dark Matter with\\  
Space-based Gravitational Wave Detectors in the mHz Regime}

\author{Jing-Rui Zhang$^{1,5,6}$}
\email[corresponding author: ]
{zhangjingrui22@mails.ucas.ac.cn}
\author{Ju Chen$^{2,6}$}
\email[corresponding author: ]
{chenju@ucas.ac.cn}
\author{Heng-Sen Jiao$^{3,7}$}
\author{Rong-Gen Cai$^{4,1,5}$}
\author{Yun-Long Zhang$^{3,1}$}
\email[corresponding author: ]
{zhangyunlong@nao.cas.cn}
\affiliation{$^{1}$School of Fundamental Physics and Mathematical Sciences,  Hangzhou   Institute for Advanced Study, UCAS, Hangzhou 310024, China.}

\affiliation{$^{2}$International Center for Theoretical Physics Asia-Pacific (ICTP-AP), University of Chinese Academy of Sciences, Beijing 100190, China}

\affiliation{$^{3}$National Astronomical Observatories, Chinese Academy of Sciences, Beijing, 100101, China}

\affiliation{$^{4}$
Institute of Fundamental Physics and Quantum Technology, Ningbo University, Ningbo 315211, China}

\affiliation{$^{5}$ Institute of Theoretical Physics, Chinese Academy of Sciences, Beijing 100190, China} 

\affiliation{$^{6}$Taiji Laboratory for Gravitational Wave Universe (Beijing/Hangzhou), University of Chinese Academy of Sciences, Beijing 100049, China}

\affiliation{$^{7}$School of Astronomy and Space Science, University of Chinese Academy of Sciences, Beijing 100049, China
}

\begin{abstract}

Spin-2 ultralight dark matter (ULDM) is a viable dark matter candidate and it can be constrained using gravitational wave (GW) observations. In this paper, we investigate the detectability of spin-2 ULDM by space-based GW interferometers. By considering a direct coupling between spin-2 ULDM and ordinary matter, we derive the corresponding response functions and sensitivity curves for various time-delay interferometry channels and calculate the optimal sensitivity curves for future millihertz GW detectors. Our results demonstrate that the space-based detectors can place stringent constraints on the coupling constant of spin-2 ULDM, reaching $\alpha \sim 10^{-10}$ around a mass of $m \sim 10^{-17} \rm eV$, surpassing current limits from ground-based detectors and pulsar timing arrays. Thus, the space-based GW detectors can serve as powerful tools not only for detecting GWs but also for probing fundamental properties of ultralight dark matter.

\end{abstract}
\maketitle

\tableofcontents
\allowdisplaybreaks

\section{Introduction}

In recent years, gravitational waves (GWs) are getting increasing attention at the frontier of both theoretical and experimental physics. Ground-based gravitational wave interferometers (GWIs) have achieved notable success in the detection of GWs, including the first discovery in 2015~\cite{LIGOScientific:2016aoc}, and the observation of GWs accompanied by electromagnetic counterparts in 2017~\cite{LIGOScientific:2017vwq}. More recently, several pulsar timing array (PTA) collaborations reported evidence for the detection of a nanohertz stochastic GW background~\cite{NANOGrav:2023gor,EPTA:2023fyk,Reardon:2023gzh,Xu:2023wog}. Also, direct and indirect methods are being used to capture evidence of primordial GWs~\cite{BICEP:2021xfz,Li:2017drr,Wang:2024gko}. The study of GWs has become a thriving field, offering a wealth of phenomena for physics and astronomy.

Apart from detecting GWs, GWIs can also be utilized to detect dark matter~\cite{Adams:2004pk,Hall:2016usm,Chen:2021apc,Heisenberg:2023urf}, particularly ultralight dark matter (ULDM). ULDM refers to ultralight bosonic particles with masses $m\lesssim1~{\rm eV}$. These particles have been extensively studied recently, as potential solutions to the problems of the cold dark matter model at small scale ~\cite{Ferreira:2020fam,Hui:2021tkt}. The oscillation of ULDM could produce signals detectable by GWIs if ULDM interacts with GWIs, such as through coupling to the test masses or laser photons. This makes it possible to directly probe ULDM with GWIs when its oscillation frequency, falls within the sensitivity range of GWIs. On the other hand, if ULDM exists around black holes, it could have non-negligible effects, such as influencing binary black hole dynamics~\cite{Aghaie:2023lan,Bromley:2023yfi,Fell:2023mtf,Mitra:2023sny} or inducing superradiance~\cite{Brito:2013wya,Brito:2015oca,Brito:2017wnc,Brito:2020lup,Dias:2023ynv}. These effects suggest that GWIs can also be used for indirect ULDM searches~\cite{Brito:2017zvb,Isi:2018pzk,Yuan:2021ebu,Yuan:2022bem,Guo:2023gfc}. Recently, the prospects of detecting ULDM with GWIs have  been widely studied ~\cite{Pierce:2018xmy,Morisaki:2018htj,Nagano:2019rbw,Grote:2019uvn,Michimura:2020vxn,Armaleo:2020efr,Nagano:2021kwx,Miller:2022wxu,Hall:2022zvi,Ismail:2022ukp,Kim:2023pkx,Yu:2023iog,Chowdhury:2023xvy,Yao:2024fie,Yu:2024enm,Arjona:2024cex}, and constraints based on the observation from current GWIs have also been placed on various types of ULDM~\cite{Guo:2019ker,Miller:2020vsl,Morisaki:2020gui,Vermeulen:2021epa,LIGOScientific:2021ffg,Aiello:2021wlp,Miller:2023kkd,Fukusumi:2023kqd,Frerick:2023xnf,Manita:2023mnc,Gottel:2024cfj,KAGRA:2024ipf,Nguyen:2024fpq}.

Depending on its spin, ULDM can be classified into different species, including spin-0, spin-1, and spin-2 particles. Considerable efforts have been made to study spin-0 and spin-1 ULDM, leading to notable constraints on their parameter space~\cite{Guo:2019ker,Miller:2020vsl,Morisaki:2020gui,LIGOScientific:2021ffg,Aiello:2021wlp,Xie:2024xex}. For spin-2 ULDM, the research on ground-based GWIs~\cite{Armaleo:2020efr,Manita:2023mnc,Guo:2023gfc} and PTAs~\cite{Armaleo:2020yml,Sun:2021yra,Unal:2022ooa,Wu:2023dnp} has been focused. The detailed investigations of spin-2 ULDM using space-based GWIs remain lacking.

Recently, space-based GWIs~\cite{LISA:2017pwj,Hu:2017mde,TianQin:2015yph} have gained increasing interest in GW astronomy. A space-based GWI consists of three separate spacecrafts, forming an approximate equilateral triangle in space. Each spacecraft acts both as a laser signal transmitter and receiver, measuring the separation between free-falling test masses inside each spacecraft. Relative changes in the separation are analyzed to determine if they originate from the effects of GWs. The data stream is processed using the time-delay interferometry (TDI) algorithm~\cite{Tinto:2004wu}, which helps eliminate major sources of noise in space-based GWIs, such as laser frequency noise and clock noise. Several space-based GWI projects have been proposed, including LISA~\cite{LISA:2017pwj}, Taiji~\cite{Hu:2017mde}, and TianQin~\cite{TianQin:2015yph}. In the future, these projects will not only provide valuable information about GWs in the millihertz band, but will also serve as detectors for searches of ULDM.

In this paper, we demonstrate the potential of space-based GWIs for detecting spin-2 ULDM. Spin-2 ULDM typically has two possible interactions with GWIs: direct coupling with standard model particles and its gravitational effect. Here, we focus on its direct coupling with ordinary matter. For its gravitational effect, one can refer to~\cite{Wu:2023dnp,Yu:2024enm}. 
The paper is organized as follows. In Section~\ref{sec2}, we revisit the basics of spin-2 ULDM. In Section~\ref{sec3}, the responses of space-based GWIs to spin-2 ULDM in different TDI combinations are calculated and compared. In Section~\ref{sec4}, possible constraints from space-based GWIs on spin-2 ULDM are presented. In Section~\ref{conclusion}, we conclude and discuss the results. Throughout the paper, we use the convention $\hbar = c = 1$.

\section{Spin-2 ULDM}\label{sec2}
The idea of spin-2 dark matter dates back to 2004~\cite{Dubovsky:2004ud}, when the massive graviton was considered as a dark matter candidate. After the bimetric theory was introduced in~\cite{Hassan:2011zd}, this idea was revitalized, with explorations of its viability as a dark matter candidate across various mass ranges~\cite{Aoki:2016zgp,Babichev:2016hir,Babichev:2016bxi,Aoki:2017cnz,Marzola:2017lbt,Aoki:2019snr,Jain:2021pnk,Manita:2022tkl,Kolb:2023dzp}. More recently, spin-2 dark matter with ultralight mass was found to be detectable with GW detectors~\cite{Armaleo:2020efr,Armaleo:2020yml,Sun:2021yra,Wu:2023dnp,Xia:2023hov,Cai:2024thd,Blas:2024jyh}. In this work, we examine the detectability of spin-2 ULDM using space-based GWIs.

For spin-2 ULDM, its effects on space-based GWIs are similar to those of GWs. Therefore, the treatment for GWs~\cite{Cornish:2001bb,Robson:2018ifk} can be adapted to account for the effects of spin-2 ULDM. The occupation number of spin-2 ULDM is as large as $\frac{\rho_{\rm DM}}{m\cdot(mv)^3}\approx 10^{71}\left(\frac{\rho_{\rm DM}}{0.4\rm GeV/cm^3}\right)\left(\frac{10^{-17}\rm eV}{m}\right)^4\left(\frac{10^{-3}}{v}\right)^3$, which makes it reasonable for us to treat it as a classical field.  

We consider the flat spacetime, such that $g_{\mn}=\eta_{\mn}$. The free action for spin-2 ULDM is given by the Pauli-Fierz action~\cite{Fierz:1939ix}
\begin{align}
S_{M}= \int {\dm}^4x\left[-\frac{1}{2} M^{\mu\nu}{\cal E}^{~~\rho\sigma}_{\mu\nu}M_{\rho\sigma} 
 - \frac{m^2}{4}(\mmn M^{\mu\nu}-M^2)\right],\label{lag}
\end{align}
where $\mmn$ is the field of spin-2 ULDM, $M=\eta^{\mn}\mmn$ is the trace of $\mmn$, and $m$ is the mass of spin-2 ULDM. ${\cal E}^{~~\rho\sigma}_{\mu\nu}$ represents the Lichnerowicz operator
\begin{align}
    {\cal E}^{~~\rho\sigma}_{\mu\nu}\equiv
-\frac{1}{2}
\bigg(&\delta^\rho_\mu \delta^\sigma_\nu \square
-\eta_{\mu\nu}\eta^{\rho\sigma}\square
+\eta^{\rho\sigma}\partial_\mu\partial_\nu 
\nonumber\\
&+\eta_{\mu\nu}\partial^\rho\partial^\sigma
-2\partial^\rho\partial_{(\mu}\delta^\sigma_{\nu)}\bigg)\ ,
\end{align}
where $\square=\partial^\mu\partial_\mu$, and $\partial_{(\mu}\delta^\sigma_{\nu)}=\frac{1}{2} (\partial_{\mu}\delta^\sigma_{\nu}+\partial_{\nu}\delta^\sigma_{\mu})$.

The equations of motion for spin-2 ULDM can then be derived from the action in Eq.~\eqref{lag}:
\begin{align}
    {\cal E}^{~~\rho\sigma}_{\mn} &M_{\rho\sigma}
    +\frac{1}{2}m^2(\mmn-{\eta}_{\mn} M)=0\ .\label{eom}
\end{align}
Consider the linearised Bianchi identities $\partial^\mu M_{\mu\nu}=\partial^\nu M$, which implies $\partial^\mu\partial^\nu M_{\mu\nu}=\Box M$, and by taking the trace of Eq.~\eqref{eom}, we obtain
\begin{align}
    M &=0\ ,\label{traceless}\\
    \partial^\mu M_{\mn} &=0\ ,\label{transverse}
\end{align}
which shows that $\mmn$ is traceless and transverse. From Eq.~\eqref{transverse}, we can see that $M_{0\nu}$ are of first order in gradients of $M_{ij}$.
The equation of motion Eq.~\eqref{eom} becomes
\begin{align}
        (\square-m^2)M_{\mn} &=0\ .\label{waveeq}
\end{align}
Thus, the field $\mmn$ obeys the wave equation.

When the observation time is shorter than the coherence time $T_{\rm coh}=2\pi/(\frac{1}{2}mv^2)\approx 8\times 10^7~{\rm s}\left(\frac{10^{-16}{\rm eV}}{m}\right)$, the spin-2 ULDM field can be written as a sum of sinusoidal wave over different polarizations~\cite{Miller:2020vsl}
\begin{align}
    \mij(t,\vec{x})= \frac{\sqrt{2\rhodm}}{\sqrt{5}m}
    \sum_A\varepsilon^{A}_{ij}
    e^{i\left(\omega t-  m\vec{v}\cdot \vec{x}{+\delta_A}\right)} \ .\label{eqMij}
\end{align}
Here, $\rhodm$ is the energy density of dark matter around the solar system, approximately $0.4~\rm GeV/cm^3$.
$\vec{v}$ is the velocity of the spin-2 ULDM, with a typical value $|\vec{v}| \sim 10^{-3}$ in the solar system, allowing us to safely approximate the energy $\omega$ of spin-2 ULDM as its mass $\omega\approx m$. 
$\delta_A$ is the random phase, and $\varepsilon^A_{ij}$ represents the polarization tensor for polarization $A$.
From Eqs.~\eqref{transverse} and \eqref{eqMij}, we can see that $M_{0\nu}$ are suppressed by the speed $v/c\sim10^{-3}$ and can be ignored~\cite{Armaleo:2020yml, Wu:2023dnp}. 

In the bimetric theory, spin-2 ULDM fields with different polarizations have the same mass $m$. The distribution of polarizations may depend on specific production mechanism of spin-2 ULDM. Here we assume the five polarization modes are equally distributed, so that an additional factor $1/\sqrt{5}$ is added. 
The spin-2 ULDM has 5 independent polarization modes~\cite{Armaleo:2020efr}:
\begin{align}
    \begin{aligned}
    \varepsilon^\times_{ij}&=\frac{1}{\sqrt{2}}(p_iq_j+q_ip_j),\ 
    \varepsilon^+_{ij}=\frac{1}{\sqrt{2}}(p_ip_j-q_iq_j),\\
    \varepsilon^{\rm L}_{ij}&=\frac{1}{\sqrt{2}}(q_ir_j+r_iq_j),\ 
    \varepsilon^{\rm R}_{ij}=\frac{1}{\sqrt{2}}(p_ir_j+r_ip_j),\\
    \varepsilon^{\rm S}_{ij}&=\frac{1}{\sqrt{6}}(3r_ir_j-\delta_{ij}),
    \end{aligned}
\end{align}
where $\hat{r}$ is the unit vector in the reference direction for polarization decomposition, $\hat{p}$ and $\hat{q}$ are two unit vectors perpendicular to $\hat{r}$ and to each other. Note that these polarization tensors are normalized such that $\varepsilon_{ij}^A \varepsilon^{A',ij} = \delta^{AA'}$, which differs from the usual conventions for GWs.

On the other hand, if the observation time exceeds the coherence time, the spin-2 ULDM field is no longer coherent and must instead be treated as a stochastic superposition of multiple plane waves with random phases, amplitudes, and velocities:
\begin{align}
    \mij(t,\vec{x})=\sum_{n=1}^N \overline{M}_{n}\varepsilon_{ij,n}e^{i\left(\omega_n t-  m\vec{v}_n\cdot \vec{x}+\delta_n\right)}.
\end{align}
Here, $N$ denotes the total number of individual waves, $\overline{M}_{n}$, $\omega_n$, $\varepsilon_{ij,n}$, $\vec{v}_n$, and $\delta_n$ represent the amplitude, energy, polarization tensor, velocity, and phase of the $n$-th wave, respectively.  
In this case, semi-coherent methods~\cite{Miller:2020vsl} can be applied to derive constraints from space-based GW interferometers.
In this work, we focus on the regime where the observation time is shorter than the coherence time.

\section{Response function for spin-2 ULDM}\label{sec3}

One of the most prominent theoretical models for spin-2 ULDM is the bimetric theory~\cite{Hassan:2011zd}. This theory does not contain any ghost degrees of freedom, and compared with dRGT massive gravity~\cite{deRham:2010ik,deRham:2010kj}, it does not depend on specific choices of fiducial tensors. In the bimetric theory, there are two independent metric tensors, with one of them coupled to the matter fields~\cite{Yamashita:2014fga}. Similar to the situation in the theory of neutrinos, initially they are not in mass eigenstates. To obtain the mass spectrum, one can linearly combine these tensors to give two mass eigenstates, with one of them massless and the other one massive. The massless metric tensor can be rescaled to recover general relativity, and the massive one can be treated as spin-2 ULDM. 

The result of the linear combination of metric tensors is that the  spin-2 ULDM couples to matter universally with a factor $\frac{\alpha}{M_{\rm Pl}}$. We can evaluate its effect by changing the frame to $\tg_{\mu\nu}={\calG}_{\mu\nu}-\frac{\alpha}{M_{\rm Pl}}M_{\mu\nu}$~\cite{Armaleo:2020yml}, where $\mpl$ is the reduced Planck mass.

Since $m\gg H$, we can ignore the Hubble constant and work with $\tg_{ij}=\delta_{ij}-\frac{\alpha}{M_{\rm Pl}}M_{ij}$. Note that this form of the metric fluctuations resembles the synchronous gauge in the treatment for GWs, in which $h_{00}=h_{0i}=0$. Therefore, when we evaluate the effect of spin-2 ULDM on GWIs, we can focus on the Shapiro effect on photons~\cite{Lee:2024oxo}, which causes a frequency shift of the photon. In this section, we calculate the response of spin-2 ULDM with space-based GWIs following a treatment similar to that of GWs ~\cite{Cornish:2001bb,Robson:2018ifk}.

\subsection{Single link}
\label{sec_res_1link}

\begin{figure}[t]
    \centering
    \includegraphics[keepaspectratio, scale=0.1]{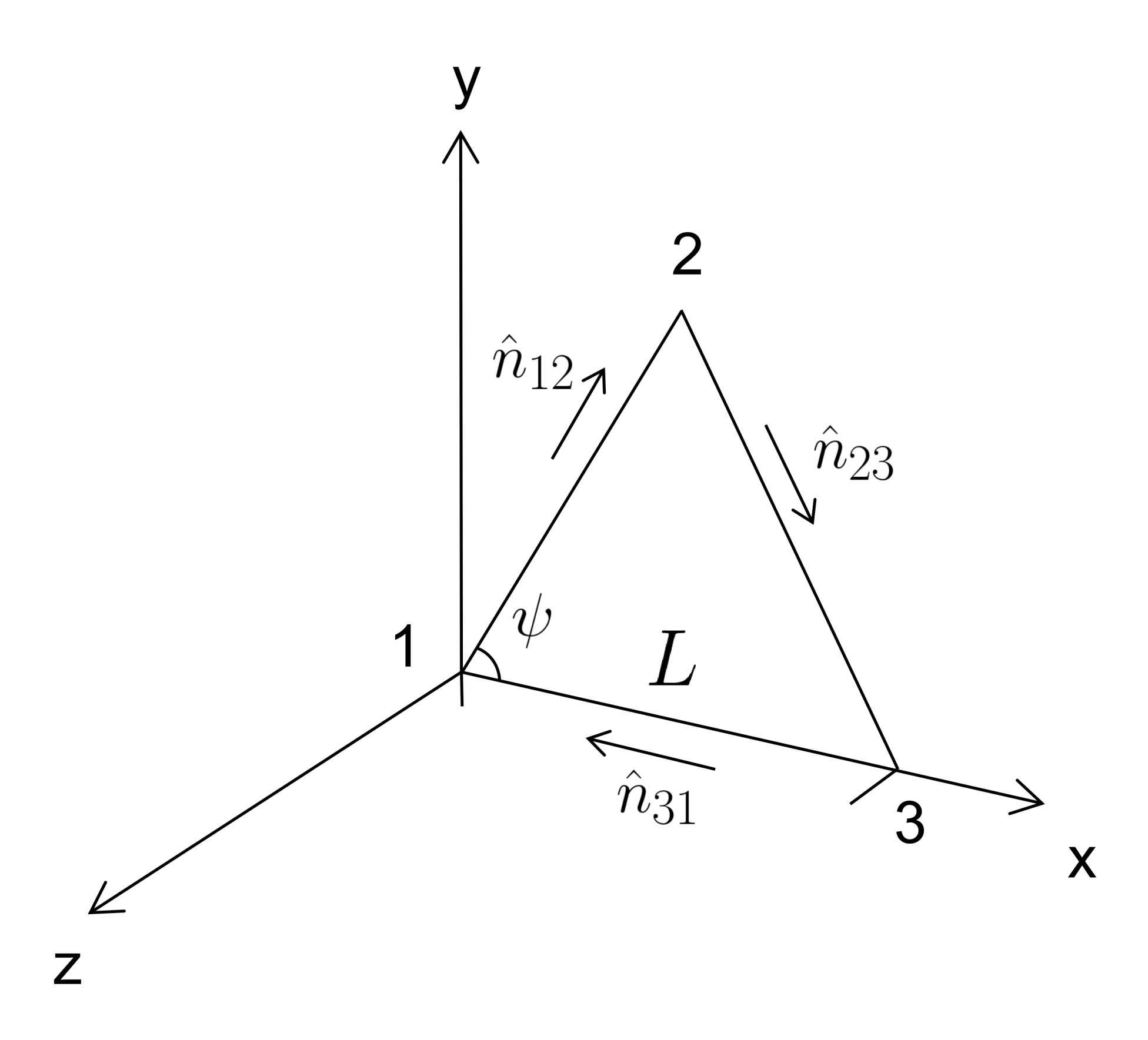}
    \caption{A schematic diagram illustrating the configuration of three detectors, where the  vertices \{1, 2, 3\} represent the distinct detectors, and $\psi = \pi/3$.}
    \label{fig_config}
\end{figure}

We first evaluate the effect of spin-2 ULDM for a single link, in which the photon simply travels from one test mass to another one. The configuration of detectors is shown in Fig. \ref{fig_config}.
For a photon with 4-momentum $p^\mu=\nu(1,-\hat{n}_{rs}^i)$, where $\hat{n}^i_{rs}$ is the unit vector pointing from the receiver $r$ to the sender $s$, the geodesic equation gives

\begin{align}
  \frac{{\dm}p^t}{{\dm}u}=\frac{\alpha\nu^2}{2M_{\rm Pl}}\partial_t M_{ij}\hat{n}_{rs}^i\hat{n}_{rs}^j\ ,
\end{align}
where $u$ is the affine parameter. After a similar treatment with GWs~\cite{Maggiore:2018sht},
we can obtain the frequency shift of the photon
\begin{align}
    z_{rs}(t)=\frac{\Delta\nu}{\nu_o}=&-\frac{\alpha}{2 M_{\rm Pl}(1+\vec{v}\cdot \hat{n}_{rs})}\hat{n}^{i}_{rs}\hat{n}^{j}_{rs}\nonumber\\ & \times\big[
    M_{ij}(t_ r,\Vec{x}_r)-M_{ij}(t_s,\Vec{x}_s)\big],
\end{align}
where $\nu_o$ is the original frequency of the photon,  $t_s$ denotes the time at which the photon is emitted by the sender $s$, $\Vec{x}_s$ represents the spatial position of the sender $s$ at $t_s$, $t_r$ denotes the time at which the photon arrives at the receiver $r$, $\Vec{x}_r$ represents the spatial position of the receiver $r$ at $t_r$.  Note that in the denominator, the factor is $(1+\vec{v}\cdot \hat{n}_{rs})$, which is different from GWs~\cite{Maggiore:2018sht}, where the factor is $(1+\hat{n}\cdot \hat{n}_{rs})$, in which $\hat{n}$ represents the direction of GWs. This is because the speed of spin-2 ULDM cannot reach the speed of light and is typically far from it. Here we adapt the usual treatment that $v/c\approx 10^{-3}$.

The frequency shift can be transformed into the change of arm length
\begin{align}
    s(t)=&\frac{\Delta L}{L}=\frac{\ell_{rs}-L}{L}=\frac{\Delta\Phi_{rs}}{\nu L}=\frac{\int_0^t z_{rs} {\dm}t}{L}\\
=&-\frac{\alpha\, \hat{n}^{i}_{rs}\hat{n}^{j}_{rs}
\big[
    M_{ij}(t+L,\Vec{x}_r)-M_{ij}(t,\Vec{x}_s)\big]
}{2i M_{\rm Pl} mL(1+\vec{v}\cdot \hat{n}_{rs})},
    \label{sOFt}
\end{align}
where $\ell_{rs}$ is the perturbed arm length and $L$ is the nominal arm length of the detector. 
Following the notation used in treatments of GWs~\cite{Cornish:2001bb}, we can view Eq.~\eqref{sOFt} as the contraction between the detector tensor $D^{ij}$ and the signal $h_{ij}$
\begin{align}
    s(t)&=\detector{} h_{ij}(t,\vec{x}_s),
\end{align}
where
\begin{align}
    \detector{}&=\frac{1}{2}\transfer\trls{rs}\nsr^i \nsr^j,
    \\
     h_{ij}(t,\vec{x}_s)&=- \frac{h_M}{\sqrt{5}}\sum_A\varepsilon^{A}_{ij}e^{i(mt-m\vec{v}\cdot \vec{x}_s +\delta_A)}. \label{h_ij}
\end{align}
Here,
\begin{equation}
h_M=\frac{\alpha\sqrt{2\rhodm}}{m\mpl},
\label{hM}
\end{equation}
and $\transfer$ is the transfer function
\begin{align}
        \transfer\trls{rs}={\rm sinc}\left[{\phim}
    \doppler{rs}\right]e^{i{\phim}\doppler{rs}},
\end{align}
where ${\rm sinc} \,x\equiv \tfrac{\sin x}{x}$, and we have introduced $\phim \equiv \frac{mL}{2}$.

We can then calculate the auto-correlation of signal $s(t)$, denoted by $\langle s^2(t)\rangle$. For spin-2 ULDM, we adopt the assumption that different polarization states are statistically uncorrelated, following a treatment similar to that used for stochastic GW background~\cite{Cornish:2001bb}. Then the auto-correlation of $s(t)$ satisfies~\cite{Cornish:2001bb}
\begin{align}
    \langle s^2(t)\rangle=S_h\response.
\end{align}
Here, $S_h$ is the power spectral density of spin-2 ULDM, and $\response$ is the response function
\begin{align}
\response=\frac{1}{5}\int \frac{{\dm}^2\hat{v}}{4\pi}\int \frac{{\dm}^2\hat{r}}{4\pi}  \sum_A \beam^{A*}(\vec{v})\beam^A(\vec{v}),
    \label{eq_response}
\end{align}
where $\int {\dm}^2\hat{r}/4\pi$ represents averaging over the directions of polarizations over the sky, and $\beam^A(\vec{v})$ is the antenna pattern function for polarization $A$
\begin{align}
\beam^A(\vec{v})=\detector{}\varepsilon^A_{ij}.
\label{eq_antenna}
\end{align}
The response function $\response$ is then averaged over all possible polarizations, resulting in a factor of $1/5$.

\begin{figure*}
    \centering
    \includegraphics[keepaspectratio, scale=0.165]{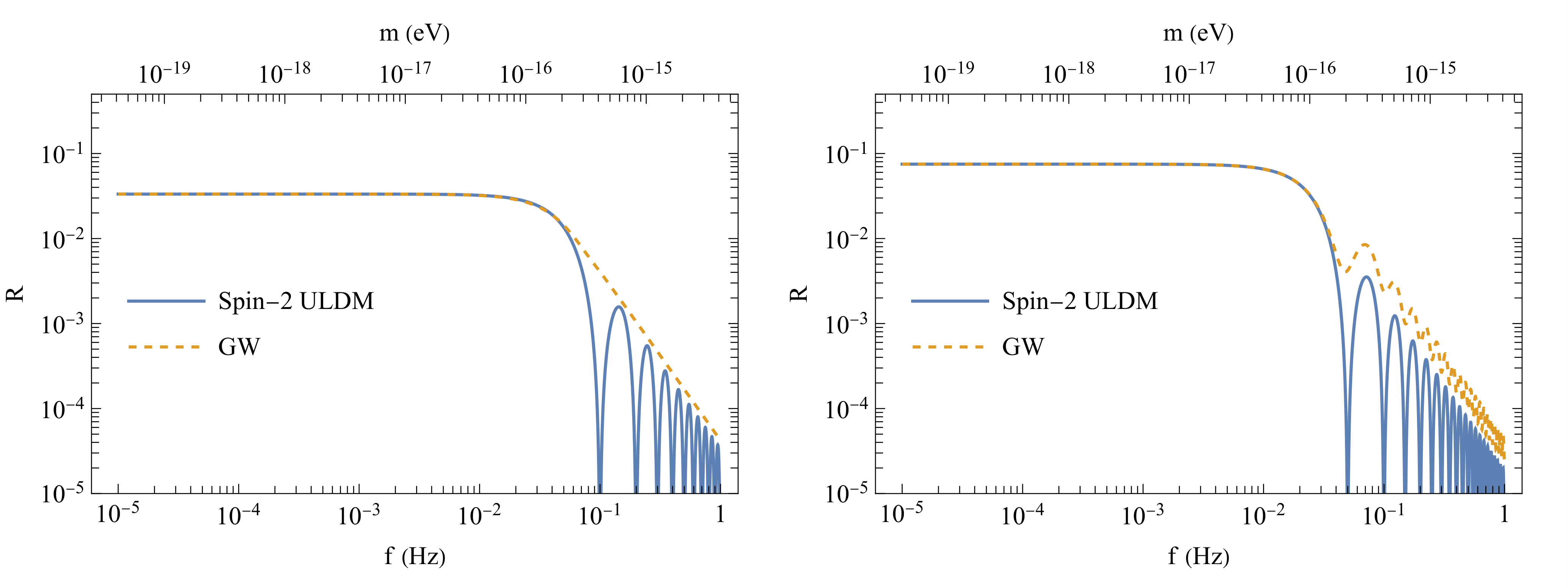}
    \caption{The response functions for single link(left) and Michelson combinations(right).}
    \label{fig_res_12}
\end{figure*}

For the single link, the response function can be analytically obtained as
\begin{align}
    \response=&
    \frac{1-v}{{60}v} {\rm sinc}^2\left[{\phim}(1-v)\right]
    -
    \frac{1+v}{{60}v} {\rm sinc}^2\left[{\phim}(1+v)\right]\nonumber\\
    &-
    \frac{{\rm Si}\left[2 \phim (1-v)\right]}{{60}v \phim }
    +
    \frac{{\rm Si}\left[2 \phim (1+v)\right]}{60v \phim },
    \label{exact_response_1link}
\end{align}
where ${\rm Si}\,x\equiv \int^x_0 \tfrac{\sin t}{t}dt$, and numerical values are calculated and shown in Fig.~\ref{fig_res_12}.
In the calculation, we assume the detector arm length to be $L=3\times 10^{9}\mathrm m$ and adopt $v=10^{-3}$, which is a typical value for the velocity of dark matter localized in the solar system. Note that the responses are expressed as functions of the mass $m$, and for comparison with GWs, we also present the corresponding frequency in the plot, given by $m = 2 \pi f$.
From the left panel of Fig.~\ref{fig_res_12}, we see that the response functions for both spin-2 ULDM and GWs approach the same constant value in the low-frequency limit. Specifically, the response function for both approaches $\frac{1}{30}$.
We also observe that for spin-2 ULDM, the response function exhibits frequent drops at higher frequencies, near integer multiples of $\frac{1}{L}$. In contrast, the response curve for GWs remains smooth, without such drops. Expanding Eq.~\eqref{exact_response_1link} around these drops, we find that the response function $\response$ becomes $\frac{v^2}{90}+{\cal O}(v^3)$, with the frequency given by $f=N(\frac{1}{L}+\frac{2 v^2}{3L})+{\cal O}(v^3)$, where $N$ is a positive integer. These drops arise from the limited velocity of spin-2 ULDM. In particular, if we assume the speed of spin-2 ULDM to be zero, the response function becomes $\response={\rm sinc}^2(mL/2)/30$, which equals zero at $f = \frac{N}{L}$. 
We have also analyzed the response function of spin-2 ULDM for different velocity directions, rather than averaging over the direction. The results remain indistinguishable from the curve in the left-hand side of the Fig.\ref{fig_res_12}, demonstrating that the direction of velocity does not contribute to the leading order of the response function. This differs from the coupling of scalar ULDM and detectors, where the velocity distribution needs to be considered. (e.g., in \cite{Roberts:2018xqn})

\subsection{Michelson configuration}
\label{sec_res_michelson}

For a normal equal-arm Michelson configuration, the signal can be expressed as:
\begin{align}
    s_{\rm m1}(t) = &\frac{1}{2L}\big[\ell_{21}(t-2L)+\ell_{12}(t-L)\nonumber\\
                    &\qquad - \ell_{31}(t-2L)-\ell_{13}(t-L)\big], \label{Michelson}
\end{align}
where $\ell_{rs}=L[1+s(t)]$, as given in Eq.\eqref{sOFt}.
Similarly, we can introduce the detector tensor of $\detector{\rm m1}$, and:
\begin{equation}
    s_{\rm m1}(t) = \detector{\rm m1} h_{ij}(t,\Vec{x}_1).
\end{equation}
With Eqs. \eqref{sOFt} and \eqref{h_ij}, we obtain
\begin{align}
    \detector{\rm m1}=\frac{1}{2}\left[n^i_{21}n^j_{21}\transfer_{\rm m1}\trls{21}-n^i_{31}n^j_{31}\transfer_{\rm m1}\trls{31}
    \right],
\end{align}
and the transfer function
\begin{align}
    \transfer_{\rm m1}\trls{rs} &=\frac{1}{2}\big[{\rm sinc}\left({\phim}\doppler{rs}\right)e^{i{\phim}(-3+ \vec{v}\cdot\hat{n}_{rs})}\nonumber\\
    &+{\rm sinc}\left({\phim}\doppleranti{rs}\right)e^{i{\phim}(-1+\vec{v}\cdot\hat{n}_{rs})}
    \big].
\end{align}
The response function can then be evaluated using Eqs.~\eqref{eq_antenna} and \eqref{eq_response}, and the result is shown in the right panel of Fig.~\ref{fig_res_12}. From the figure, we can see that the general behavior of spin-2 ULDM resembles GWs, with two main differences. Firstly, the low-frequency limit of the response function is 0.075.
Note that we use the normalization convention $\varepsilon_{ij}^A \varepsilon^{A', ij} = \delta^{AA'}$, and have averaged over all polarizations, so the value here is different from the value found in some other literature~\cite{Maggiore:2018sht}. Secondly, the drop at high frequencies is more pronounced for spin-2 ULDM compared to GWs, due to the relatively slow speed of spin-2 ULDM, as analyzed in Sec.~\ref{sec_res_1link}.  Regarding the ULDM velocity direction, similarly as in Sec. \ref{sec_res_1link}, the response functions for the Michelson configuration remain indistinguishable across different velocity orientations.

\subsection{Michelson TDI combinations}

For space-based GWIs, the distances between different arms cannot be strictly fixed to be equal. As a result, laser frequency noise becomes dominant and can far exceed GW signals. To address this problem, time-delay interferometry (TDI) was introduced, enabling virtual equal-arm interference through a linear combination of time-shifted signals~\cite{Tinto:2004wu}. We calculate and compare the response of common TDI channels. Since the final sensitivity is independent of the TDI generation, provided that the laser noise is sufficiently suppressed, we consider only the first-generation TDI (or TDI 1.5) as a demonstration. For simplicity, we assume that all arm lengths are constant and equal.

Here, we present the results for the Michelson TDI configurations. The formulas for the Sagnac and AET combinations are provided in Appendix~\ref{sec:Sagnac} and Appendix~\ref{sec:AET}, respectively.
For Michelson X combinations, the signal is
\begin{align}
    s_{\rm X}(t)=&\frac{1}{4L}\big[\ell_{21}(t-4L)+\ell_{12}(t-3L)+\ell_{31}(t-2L)\nonumber\\&~+\ell_{13}(t-L)\nonumber-\ell_{31}(t-4L)-\ell_{13}(t-3L)\nonumber\\&~-\ell_{21}(t-2L)-\ell_{12}(t-L)\big]\label{eq_MX}\\=&\detector{\rm X} h_{ij}(t,\vec{x}_1).
\end{align}
The corresponding detector tensor is given by
\begin{align}
    \detector{\rm X}=\frac{1}{2}\left[n^i_{21}n^j_{21}\transfer_{\rm mx}\trls{21}-n^i_{31}n^j_{31}\transfer_{\rm mx}\trls{31}
    \right],
\end{align}
and the transfer function
\begin{align}
    \transfer_{\rm mx}\trls{rs}=-\frac{1}{2}(1-e^{-4i\phim})\transfer_{\rm m1}\trls{rs}.
\end{align}
The Michelson Y and Z combinations can be obtained by cyclic permutations of the spacecraft labels \{1, 2, 3\} in the expression of $X$. After averaging the direction of velocity, the response functions of X, Y and Z channels for spin-2 ULDM become the same, which is confirmed by our numerical calculations.

\begin{figure*}[htbp]
    \centering
    \includegraphics[keepaspectratio, scale=0.165]{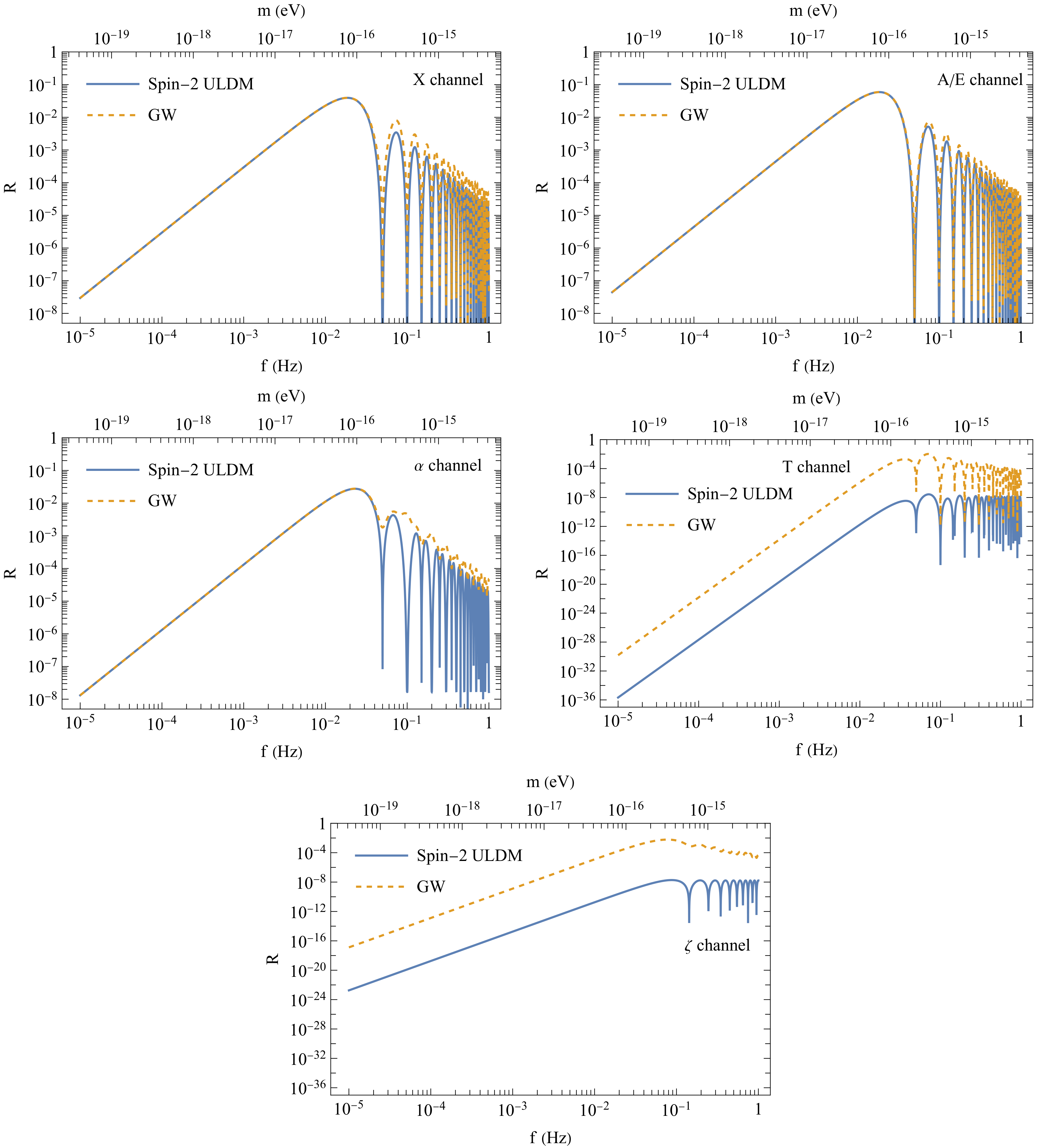}
    \caption{The response functions for different TDI channels.}
    \label{fig_res_tdi}
\end{figure*}

The response functions of spin-2 ULDM are shown in Fig.~\ref{fig_res_tdi}, which are compared with those of GWs. The response functions of GWs are calculated in a manner analogous to those for spin-2 ULDM, with two main differences. Firstly, the polarization tensors for GWs are restricted to ``+'' and ``$\times$'' modes. Secondly, the velocity $\vec{v}$ and the reference vector $\hat{r}$ for spin-2 ULDM are replaced by the directions of GWs, denoted by $\hat{n}$, and averaging over $\hat{v}$ and $\hat{r}$ is substituted by averaging over $\hat{n}$ in Eq. \eqref{eq_response}. In the low-frequency limit, the response function behaves as $\frac{3}{10}\pi^2L^2f^2$. At relatively high frequencies, both responses exhibit similar behavior. 
This can be understood by comparing the Michelson configuration in Sec.~\ref{sec_res_michelson} with the X channel here. The response functions of the two channels differ by a factor of $\sin^2(mL)$, which drives the response function of GWs to zero at the drops in high frequency, making its response curve more closely resemble that of spin-2 ULDM.

In general, the low ULDM velocity $v/c$ adds a tiny perturbation to the response function, which is negligible. However, this perturbation has relatively considerable effects. 
To be specific, for symmetric channels, such as Sagnac $\zeta$ and T channels, the whole response function is zero when the ULDM velocity is set to zero. The non-vanishing ULDM velocity is indispensable for the nonzero response function in these channels. 
For channels which are not symmetric over the three detectors, such as single link, Michelson combinations and Sagnac $\alpha$, the velocity $v$ contributes to nonzero values when the response function reaches its local minima. If the velocity of ULDM is set to zero, the response function at these local minima becomes zero. Consistent with the findings in Sec. \ref{sec_res_michelson}, the response functions for the Michelson X channel exhibit no measurable variation with respect to ULDM velocity direction.

\begin{figure*}
\centering
\centerline{\includegraphics[keepaspectratio, scale=0.165]{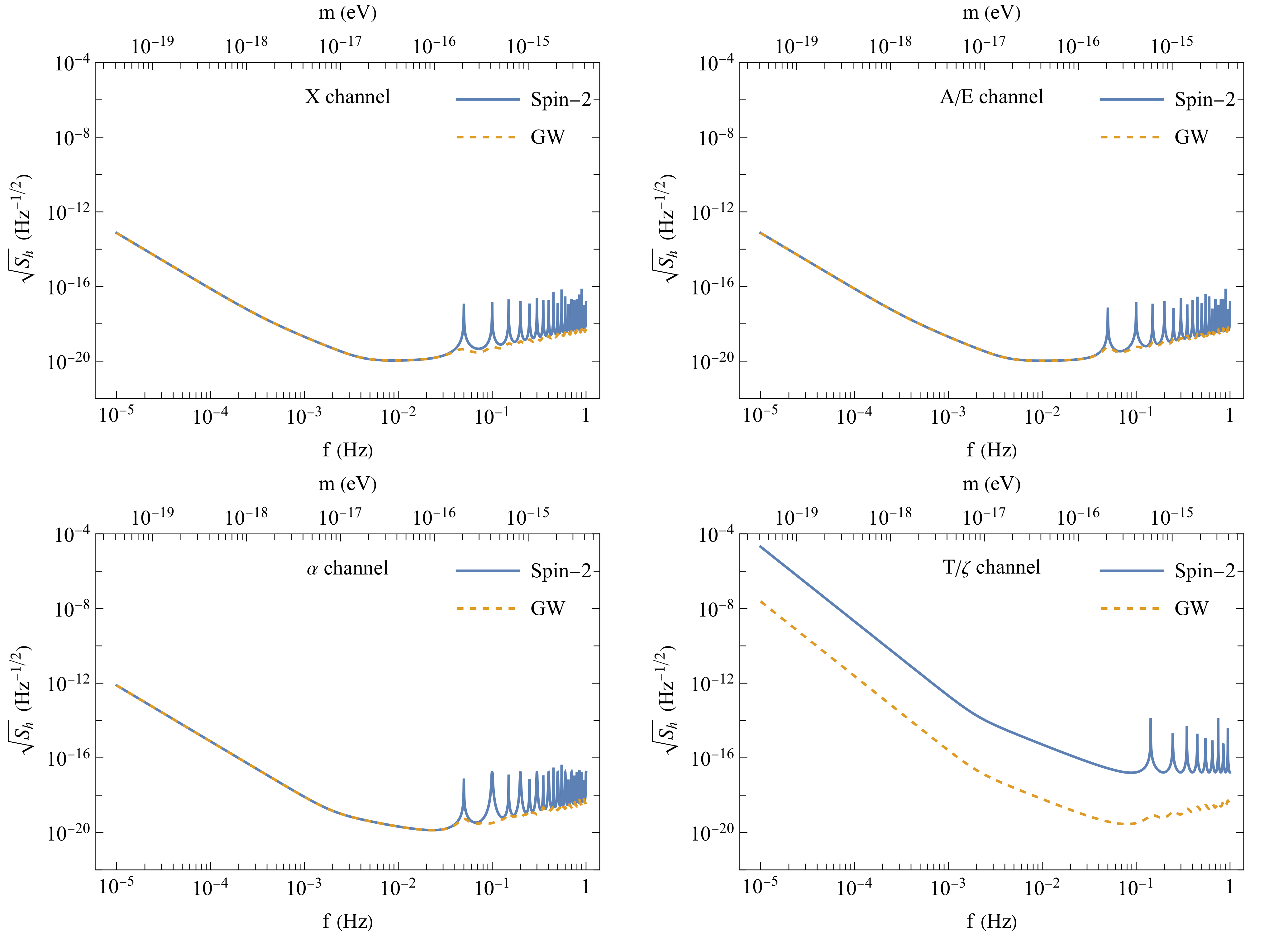}}
\caption{The sensitivity curves of GWIs for the detection of spin-2 ULDM in different TDI channels, compared with the performance of GWs. We assume $L=3\times 10^9~{\mathrm m}$ and $v=10^{-3}$.} 
\label{fig_sen_4fig}
\end{figure*}

\section{Constraints on spin-2 ULDM}\label{sec4}

\begin{figure}[h]
\centering
\centerline{\includegraphics[keepaspectratio, scale=0.119]{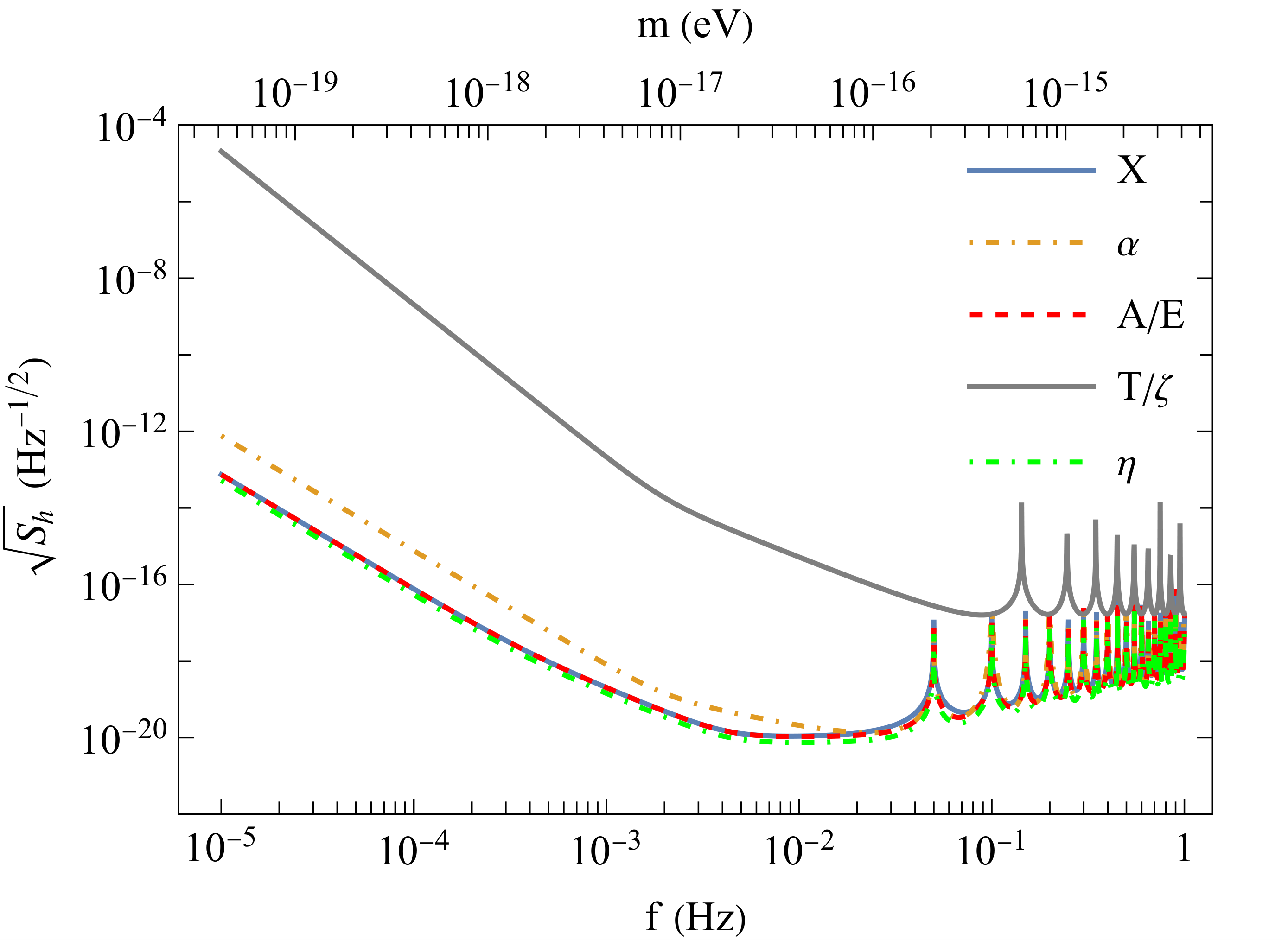}}
\caption{The comparison between sensitivity curves of different channels for spin-2 ULDM.} 
\label{fig_sen_4channel}
\end{figure}

\begin{figure}[h]
\centering
\centerline{\includegraphics[keepaspectratio, scale=0.119]{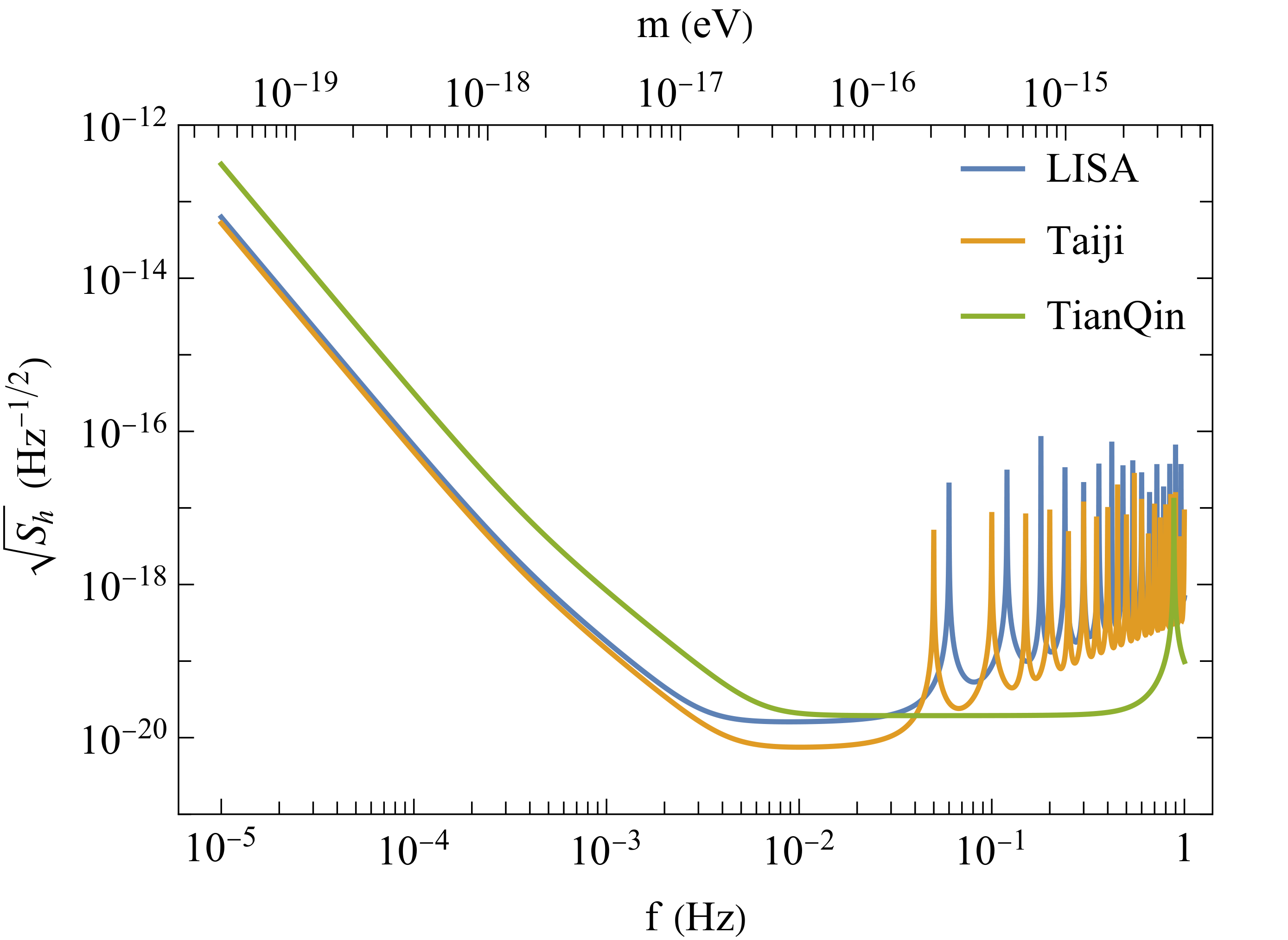}}
\caption{The comparison between optimal sensitivity curves of different detectors for spin-2 ULDM.} 
\label{fig_sen_opt}
\end{figure}

In order to quantitatively examine the detectability of space-based GW detectors on spin-2 ULDM, we calculate the sensitivity in different TDI channels and then give the constraint on the coupling constant $\alpha$ for spin-2 ULDM with different mass $m$.

\subsection{Sensitivity curves of space-based GWIs} 

The sensitivity of each TDI channel can be obtained through
\begin{align}
    S_{h}=\frac{N_h}{\response_h},
\end{align}
where $h$ is the label for each channel. The response functions $\response_h$ for spin-2 ULDM and GWs are calculated in Sec.~\ref{sec3}. $N_h$ denotes the one-sided noise spectral density. The remaining secondary noises after the TDI process include the Optical Metrology System (OMS) noise and the test mass acceleration (acc) noise. For LISA~\cite{LISA:2017pwj}, Taiji~\cite{Hu:2017mde} and TianQin~\cite{TianQin:2015yph}, the power spectral densities of these noises are given by
\begin{align}
P_{\rm OMS}=&A^2_{\rm OMS}\bigg[1+\Big(\frac{2  {\rm mHz}}{f}
    \Big)^4 \bigg]\,,\\
P_{\rm acc} =&A^2_{\rm acc}\bigg[1+\Big(\frac{0.4{\rm mHz}}{f}
    \Big)^2  \bigg] 
\bigg[1+\Big(\frac{f}{8  {\rm mHz}}
    \Big)^4  \bigg],
\end{align}
where $A_{\rm OMS}$, $A_{\rm acc}$ and arm length $L$ for different detectors are listed in Table \ref{table_detectors}.
For different TDI channels, we utilize the noise spectral density derived in~\cite{Yu:2023iog}, with modification to adjust to our convention:
\begin{align}
    N_{\rm X}&=\frac{1}{L^2}\sin^2\frac{f}{f_*}\left[P_{\rm OMS}+\big(3+\cos\frac{2f}{f_*}\big)\frac{P_{\rm acc}}{(2\pi f)^4}
    \right],
    \nonumber\\
    N_{\rm A}&=N_{\rm E}=\frac{1}{2L^2}\sin^2\frac{f}{f_*}\big[\big(2+\cos\frac{f}{f_*}\big)P_{\rm OMS}\nonumber\\
    &\ \ \ \ \ \ \ \ \ \ \ \ \ +\big(6+4\cos\frac{f}{f_*}+2\cos\frac{2f}{f_*}\big)\frac{P_{\rm acc}}{(2\pi f)^4}
    \big],
   \nonumber \\
    N_{\rm T}&=\frac{2}{L^2}\sin^2\frac{f}{2f_*}\sin^2\frac{f}{f_*}\big[P_{\rm OMS}+4\sin^2\frac{f}{2f_*}\frac{P_{\rm acc}}{(2\pi f)^4}
    \big],
    \nonumber\\
    N_{\rm \alpha}&=\frac{1}{9L^2}\big[6P_{\rm OMS}+\big(16\sin^2\frac{f}{2f_*}+8\sin^2\frac{3f}{2f_*}
    \big)\frac{P_{\rm acc}}{(2\pi f)^4}
    \big],
   \nonumber \\
    N_{\rm \zeta}&=\frac{2}{3L^2}\bigg[P_{\rm OMS}+4\sin^2\frac{f}{2f_*}\frac{P_{\rm acc}}{(2\pi f)^4}
    \bigg],
\end{align}
where $f_*\equiv c/(2\pi L)$.

\begin{table}[ht]
\centering
\renewcommand{\arraystretch}{1.5}
\begin{tabular}{|m{1.2cm}<{\centering}|m{2cm}<{\centering}|m{2cm}<{\centering}|m{1.4cm}<{\centering}|}
\hline
 & $A_{\rm OMS}$ & $A_{\rm acc}$ & $L$  \\ \hline
LISA & $15 {\rm pm}/\sqrt{{\rm Hz}}$ & $3{\rm fm/s^2}/\sqrt{{\rm Hz}}$ &  $2.5 {\rm Gm}$  \\ \hline
Taiji  & $8{\rm pm}/\sqrt{{\rm Hz}}$  &  $3{\rm fm/s^2}/\sqrt{{\rm Hz}}$ &  $3.0 {\rm Gm}$  \\ \hline
TianQin  &  $1{\rm pm}/\sqrt{{\rm Hz}}$ & $1{\rm fm/s^2}/\sqrt{{\rm Hz}}$  &  $0.17 {\rm Gm}$  \\ \hline
\end{tabular}
\caption{Noise amplitude spectral density parameters and arm lengths for different space-based GW detectors.}
\label{table_detectors}
\end{table}

The sensitivity curves of spin-2 ULDM for different channels, along with those of GWs, are shown in Fig.~\ref{fig_sen_4fig}. The comparison between different channels is depicted in Fig.~\ref{fig_sen_4channel}. As shown in the plots, the symmetric channels, including the fully symmetric Sagnac combination $\zeta$ and channel T, exhibit a much worse sensitivity to spin-2 ULDM, being three orders of magnitude higher than those of GWs. Combined with the results in~\cite{Yu:2023iog}, these channels may serve as promising candidates for mitigating the effects of ULDM on GWIs during gravitational wave detection, particularly at higher frequencies when $f\gtrsim 10 {\rm mHz}$, where the sensitivity of GWs in symmetric channels is comparable to that of other asymmetric TDI channels. The sensitivities of the A/E and X channels are similar at low frequencies, while at higher frequencies the A/E channels show slightly better sensitivity. For the Sagnac $\alpha$ channel, the sensitivity is also low at low frequencies, but at higher frequencies, it becomes comparable to that of the A channel. 
We also calculate the optimal sensitivity curve, which combines the sensitivities of all independent TDI channels~\cite{Prince:2002hp,Yu:2023iog},
\begin{equation}
    1/S_\eta = 1/S_{\rm A} + 1/S_{\rm E} + 1/S_{\rm T},
\end{equation}
and use it to place constraints on spin-2 ULDM. The result is shown in Fig.~\ref{fig_sen_opt}, where the optimal sensitivity curves for three different space-based GWIs are presented. We can observe that LISA and Taiji exhibit better sensitivity at low frequencies, while TianQin performs better at higher frequencies due to its shorter arm length.

\subsection{Constraints on spin-2 ULDM}
\label{sec5}

For spin-2 ultralight dark matter, there are two main parameters: the mass $m$ and the coupling constant $\alpha$. The mass corresponds to the frequency of the signal that appears in the detector response. The coupling constant describes the potential nontrivial interactions between spin-2 ULDM and matter 
as outlined in bimetric theory~\cite{Marzola:2017lbt}. Using the optimal sensitivity introduced in the previous section, we can place constraints on $\alpha$ for each mass of spin-2 ULDM.
As the observation time $T_{\rm obs}$ increases, the signal would accumulate as
\begin{align}
    h_M\sqrt{T_{\rm obs}}=
\frac{\alpha\sqrt{2\rhodm}}{m\mpl}\sqrt{T_{\rm obs}},
\end{align}
where $h_M$ is the amplitude of the signal $h_{ij}$ in \eqref{hM}. 
When the accumulated signal reaches the sensitivity of detectors $\sqrt{S_\eta}$, it can eventually be detected~\cite{Yu:2023iog}.

\begin{figure}[h]
\centering
\centerline{\includegraphics[keepaspectratio, scale=0.61]{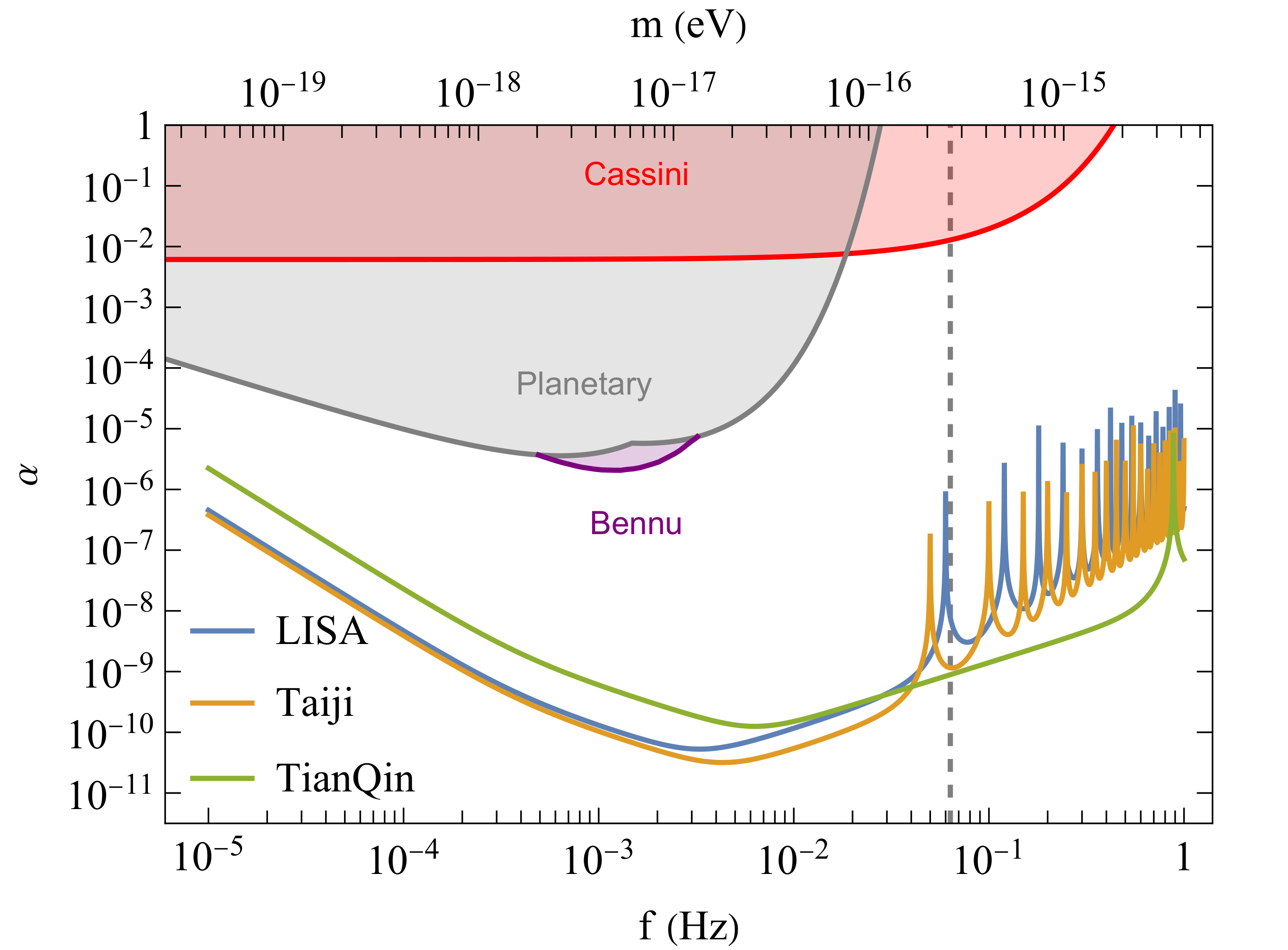}}
\caption{The constraints on coupling parameter $\alpha$ for spin-2 ULDM. The observation time is set to be $T_{\rm obs}=1~{\rm yr}$. 
(To the right of the gray dashed line, the coherence time of spin-2 ULDM is shorter than $T_{\rm obs}$, the data need to be analyzed in discrete segments and curves in this regime are only for illustration.)
Other known constraints are also shown for comparison. The red shaded area is excluded by solar system tests~\cite{Hohmann:2017uxe}. The grey shaded area is excluded by planetary motions~\cite{Sereno:2006mw}. The purple shaded area is excluded by asteroid observations~\cite{Tsai:2023zza}. }
\label{fig_constraint}
\end{figure}

Here we assume the observation time is shorter than the coherence time of spin-2 ULDM, 
then the approximation that spin-2 ULDM can be treated as monochromatic plane waves still holds.
  With this, we can get the constraint on $\alpha$ for different mass parameters:
\begin{align}
\alpha= \frac{m\mpl}{\sqrt{T_{\rm obs}}}\sqrt{\frac{S_\eta}{2\rhodm}}.
\end{align}
The final constraints are shown in Fig.~\ref{fig_constraint}. We find that the strongest constraint on $\alpha$ is about $10^{-10}$, achieved by Taiji at $m\approx10^{-17}{\rm eV}$, which is a very strong limit compared to current results from ground-based GWIs and PTAs. Additionally, LISA and Taiji exhibit better sensitivity at low frequencies in the search for spin-2 ULDM, while TianQin shows improved sensitivity at frequencies $f\gtrsim 0.04 ~\rm Hz$. It is worth noting that if spin-2 ULDM constitutes only a fraction of the total dark matter density, the constraint is actually on $\sqrt{\rho_{\rm 2}/\rho_{\rm DM}} \alpha$ instead of $\alpha$, where  $\rho_{\rm 2}/\rho_{\rm DM}$ represents the fraction of spin-2 ULDM in the total dark matter.

In Fig.~\ref{fig_constraint}, some other known constraints are also shown for comparison. The red shaded area is excluded by solar system tests~\cite{Hohmann:2017uxe}. The grey shaded area is excluded by planetary motions~\cite{Sereno:2006mw}. The purple shaded area is excluded by asteroid observations~\cite{Tsai:2023zza}.
Besides, it has been pointed out that the observation for the motion of binary pulsars can be utilized to constrain spin-2 ULDM \cite{Armaleo:2019gil}.

\section{Conclusion}
\label{conclusion}
In this paper, we investigate the potential of space-based GWIs for detecting spin-2 ULDM. We examine the effect of spin-2 ULDM on the detector, considering the direct coupling between spin-2 ULDM and ordinary matter. We calculate the corresponding response functions and sensitivity curves for various TDI channels. Our results show that future space-based GWIs could place stringent constraints on the coupling constant of spin-2 ULDM, reaching $10^{-10}$ at $m \approx 10^{-17}~\rm eV$, which is about two orders of magnitude stronger than current results from ground-based GWIs and PTAs.
It is also interesting that fully symmetric TDI channels exhibit worse sensitivity to spin-2 ULDM compared to GWs. This difference may provide an opportunity to distinguish ULDM signals from GWs when sufficient observational data are accumulated.

When the observation time is shorter than the coherence time of ULDM, a fully coherent method can be utilized for ULDM detection, such as matched filtering methods~\cite{Miller:2023kkd}. Note that since the signal power is constrained to one frequency bin, the optimal filter for ULDM at each frequency is simply a monochromatic signal. On the other hand, when the observation time exceeds ULDM's coherence time, fully coherent methods become inapplicable. This is because spin-2 ULDM's frequency variations cause the signal power to spread across multiple frequency bins. Consequently, a semi-coherent method~\cite{Miller:2020vsl,Miller:2023kkd} was developed to address this challenge. In this approach, the total observation time is partitioned into shorter segments, each selected such that ULDM signals remain confined to a single frequency bin after Fast Fourier Transform (FFT). These ULDM signals are then combined incoherently to enhance detection sensitivity.

For monochromatic GWs, such as those emitted by binary white dwarfs, these continuous GWs emitted by binary systems exhibit frequency evolution over time. In contrast, ULDM signals show stochastic frequency fluctuations around $f=\frac{m}{2\pi}$.  With sufficient data accumulation, these distinct frequency behaviors enable ULDM signals to be distinguished from monochromatic GWs.
What's more, GW signals exhibit 2 polarizations while spin-2 ULDM has 5 polarization states.
On the other hand, for purely monochromatic sources without frequency modulation, such as narrow noise lines, the frequency modulation induced by the velocity distribution of ULDM offers a way to distinguish ULDM signals from these sources. For example, it has been proposed that~\cite{Miller:2020vsl}, by increasing the FFT time, the critical ratio of ULDM signal peaks exhibits a characteristic curve compared with purely monochromatic signals, which can help us recognize ULDM signals.

It is worth noting that we assume the speed of spin-2 ULDM is fixed at $10^{-3}$ in our calculations. In reality, the speed of virialized ULDM follows a Maxwellian distribution, and the motion of the solar system should also be taken into consideration. Stochastic fluctuations of spin-2 ULDM are also important factors~\cite{Yao:2024fie}. After careful consideration of these effects, the constraint for the coupling constant may be weaken. In the future, we would conduct a more comprehensive study on the detectability of spin-2 ULDM with space-based GWIs.

\begin{acknowledgments}
This work is supported by 
the National Key Research and Development Program of China (Nos. 2023YFC2206200, 2020YFC2201502, 2021YFA0718304), 
the National Natural Science Foundation of China (No.12375059 and No.12235019), the Fundamental Research Funds for the Central Universities (No. E2ET0209X2), and the Project of National Astronomical Observatories, Chinese Academy of Sciences (No. E4TG6601).
We thank Yong Tang and Jiang-Chuan Yu for many helpful discussions.
\end{acknowledgments}

\appendix

\section{Sagnac TDI combinations}
In this section, we summarise the detector tensor and transfer functions in Sagnac TDI combinations.

\subsection{Sagnac combinations}
\label{sec:Sagnac}

For Sagnac combination $\alpha$,
the signal
\begin{align}
    s_{\alpha}(t)=&\frac{1}{3L}\big[\ell_{31}(t-3L)+\ell_{23}(t-2L)+\ell_{12}(t-L)\nonumber\\
    &-\ell_{21}(t-3L)-\ell_{32}(t-2L)-\ell_{13}(t-L)\big]
    \label{eq_sa}
    \nonumber\\
    =&\detector{\alpha} h_{ij}(t,\Vec{x}_1).
\end{align}
The corresponding detector tensor is given by
\begin{align}
    \detector{\alpha}=&\frac{1}{6}\big[n^i_{21}n^j_{21}\transfer_{\alpha,21}\trls{21}+ n^i_{32}n^j_{32}\transfer_{\alpha,32}\trls{32}\nonumber\\
    &+n^i_{13}n^j_{13}\transfer_{\alpha,13}\trls{13}
    \big],
\end{align}
where the transfer functions are
\begin{align}
    \transfer_{\alpha,21}\trls{21}=&-{\rm sinc}\left({\phim}\doppler{21}\right)e^{i{\phim}(-5+ \vec{v}\cdot\hat{n}_{21})}\nonumber\\
    &+{\rm sinc}\left({\phim}\doppleranti{21}\right)e^{i{\phim}(-1+\vec{v}\cdot\hat{n}_{21})}
    ,\nonumber\\
    \transfer_{\alpha,32}\trls{32}=&\big[-{\rm sinc}\left({\phim}\doppler{32}\right)\nonumber\\
    +{\rm sinc}&\left({\phim}\doppleranti{32}\right)
    \big]
    e^{i{\phim}(-3+\vec{v}\cdot(\hat{n}_{21}-\hat{n}_{13}))}
    ,\nonumber\\
    \transfer_{\alpha,13}\trls{13}=&-{\rm sinc}\left({\phim}\doppler{13}\right)e^{i{\phim}(-1- \vec{v}\cdot\hat{n}_{13})}\nonumber\\
    &+{\rm sinc}\left({\phim}\doppleranti{13}\right)e^{i{\phim}(-5-\vec{v}\cdot\hat{n}_{13})}.
\end{align}
Note that the transfer functions for different arms here have distinct forms, which is due to our choice of $h_{ij}(t,\vec{x}_1)$. The Sagnac combinations $\beta$ and $\gamma$ are cyclic permutations of Eq.~\eqref{eq_sa}.

The comparison between spin-2 ULDM and GWs for Sagnac $\alpha$ combinations is depicted in Fig.~\ref{fig_res_tdi}. At low frequencies, the low frequency limit of the response function is $\frac{2}{15}\pi^2L^2f^2$. At high frequencies, the spin-2 ULDM exhibits different behavior compared to GWs. When the frequency is near odd multiples of $\frac{1}{2L}$, both response curves drop, with the response for spin-2 ULDM being lower due to its lower speed. When the frequency is near even multiples of $\frac{1}{2L}$, the response for spin-2 ULDM drops, while the response for GWs remains unaffected. As for the direction of ULDM velocity, the response function for the Sagnac $\alpha$ channel also shows insensitivity with different directions, which is similar with the case in \ref{sec_res_michelson}.

\subsection{Fully symmetric Sagnac combination}
\label{sec_res_ss}
The fully symmetric Sagnac combination $\zeta$ is given by
\begin{align}
    s_{\zeta}(t)=&\frac{1}{3L}\big[\ell_{21}(t-L)-\ell_{12}(t-L)+\ell_{13}(t-L)\nonumber\\
    &-\ell_{31}(t-L)+\ell_{32}(t-L)-\ell_{23}(t-L)\big]
    \label{eq_ss}
\nonumber\\
=&\detector{\zeta} h_{ij}(t,\Vec{x}_1).
\end{align}
The corresponding detector tensor is
\begin{align}
    \detector{\zeta}=\frac{1}{6} \big[&n^i_{21}n^j_{21}\transfer_{\zeta,21}\trls{21}+ n^i_{32}n^j_{32}\transfer_{\zeta,32}\trls{32}\nonumber\\
    &+ n^i_{13}n^j_{13} \transfer_{\zeta,13}\trls{13}
    \big]
    \big],
\end{align}
where the transfer functions are
\begin{align}
&\transfer_{\zeta,21} \trls{21}=e^{i{\phim}(-1+ \vec{v}\cdot\hat{n}_{21})}
   \nonumber\\
&~\times \big[
    {\rm sinc}\left({\phim}\doppler{21}\right)-{\rm sinc}\left({\phim}\doppleranti{21}\right)\big] , \nonumber\\
&\transfer_{\zeta,32} \trls{32}=e^{i{\phim}(-1+ \vec{v}\cdot(\hat{n}_{21}-\hat{n}_{13}))}
  \nonumber\\
&~\times \big[
    {\rm sinc}\left({\phim}\doppler{32}\right)  -{\rm sinc}\left({\phim}\doppleranti{32}\right)\big] , \nonumber\\
&\transfer_{\zeta,13} \trls{13}=e^{i{\phim}(-1- \vec{v}\cdot\hat{n}_{13})}
   \nonumber\\
&~\times\big[
    {\rm sinc}\left({\phim}\doppler{13}\right)  -{\rm sinc}\left({\phim}\doppleranti{13}\right)\big] .
\end{align}
As shown in Fig.~\ref{fig_res_tdi}, it is obvious that the response of GWs greatly exceeds that of spin-2 GM for the $\zeta$ channel. This can be seen by Taylor expanding the response function for spin-2 ULDM with velocity $v$. The result is 
\begin{align}
    \response_{\zeta}=\frac{\left(2\sin{\phim}-2\phim \cos{\phim}\right)^2}{240\phim^2} v^2 +{\cal O}(v^4),
    \label{eq_res_ss_taylor_v}
\end{align} 
which has no zeroth-order terms. Therefore, the response function is strongly suppressed by the velocity $v$. In the symmetric channel 
$\zeta$, the response function exhibits a directional dependence on the ULDM velocity. When the ULDM velocity is perpendicular to the detector plane, the response function vanishes entirely. Conversely, when the ULDM velocity aligns parallel to the detector plane, the response function reaches its maximum value, which is 50\% larger than its direction-averaged counterpart.

\section{AET channels}
\label{sec:AET}

For AET channels~\cite{Prince:2002hp}, the signals are given by
\begin{align}
    s_{\rm A}(t)&=\frac{1}{\sqrt{2}}[s_{\rm Z}(t)-s_{\rm X}(t)]=\detector{\rm A} h_{ij}(t,\Vec{x}_1),\nonumber\\
    s_{\rm E}(t)&=\frac{1}{\sqrt{6}}[s_{\rm X}(t)-2s_{\rm Y}(t)+s_{\rm Z}(t)]=\detector{\rm E} h_{ij}(t,\Vec{x}_1),\nonumber\\
    s_{\rm T}(t)&=\frac{1}{\sqrt{3}}[s_{\rm X}(t)+s_{\rm Y}(t)+s_{\rm Z}(t)]=\detector{\rm T} h_{ij}(t,\Vec{x}_1).
\end{align}
The relating functions are calculated as follows. The detector tensor for A channel is given by
\begin{align}
    \detector{\rm A}=\frac{1}{4\sqrt{2}} \big[&n^i_{21}n^j_{21}\transfer_{\rm A,21}\trls{21}+ n^i_{32}n^j_{32}\transfer_{\rm A,32}\trls{32}\nonumber\\
    &+n^i_{13}n^j_{13}\transfer_{\rm A,13}\trls{13}
    \big]
    \big],
\end{align}
where the transfer functions are
\begin{align}
&\transfer_{\rm A,21}\trls{21}=
    ~i\sin(2\phim)\big[
    {\rm sinc}\left({\phim}\doppler{21}\right)\nonumber\\
&\quad +{\rm sinc}\left({\phim}\doppleranti{21}\right)e^{i2\phim}\big] e^{i{\phim}(-5+ \vec{v}\cdot\hat{n}_{21})},
   \nonumber \\
&\transfer_{\rm A,32}\trls{32}=  ~i\sin(2\phim)\big[{\rm sinc}\left({\phim}\doppler{32}\right)e^{i2\phim}\nonumber\\
&\quad+{\rm sinc}\left({\phim}\doppleranti{32}\right)
    \big]
    e^{i{\phim}(-5+\vec{v}\cdot(\hat{n}_{21}-\hat{n}_{13}))},
    \nonumber\\
&\transfer_{\rm A,13}\trls{13}= -2i\sin(2\phim)\big[{\rm sinc}\left({\phim}\doppler{13}\right)\nonumber\\
&\quad+{\rm sinc} \left({\phim}\doppleranti{13}\right)
    \big]
    \cos({\phim})e^{i{\phim}(-4-\vec{v}\cdot\hat{n}_{13})}.
\end{align}

The detector tensor for E channel is calculated as
\begin{align}
    \detector{\rm E}=\frac{1}{4\sqrt{6}} \big[&n^i_{21}n^j_{21}\transfer_{\rm E,21}\trls{21}+ n^i_{32}n^j_{32}\transfer_{\rm E,32}\trls{32}\nonumber\\
    &+n^i_{13}n^j_{13}\transfer_{\rm E,13}\trls{13}
    \big]
    \big],
\end{align}
where the transfer functions are
\begin{align}
&\transfer_{\rm E,21}\trls{21}=
    -i\big[
    \left(1+2e^{i2\phim}\right){\rm sinc}\left({\phim}\doppler{21}\right)\nonumber\\
&\qquad +\left(2+e^{i2\phim}\right){\rm sinc}\left({\phim}\doppleranti{21}\right)\big]\nonumber \\
&\qquad\times
    \sin(2\phim) e^{i{\phim}(-5+ \vec{v}\cdot\hat{n}_{21})},
\nonumber\\
&\transfer_{\rm E,32}\trls{32}=
    i\big[
    \left(2+e^{i2\phim}\right){\rm sinc}\left({\phim}\doppler{32}\right)\nonumber\\
&\qquad+\left(1+2e^{i2\phim}\right){\rm sinc}\left({\phim}\doppleranti{32}\right)\big] \nonumber\\
&\qquad\times \sin(2\phim)e^{i{\phim}(-5+ \vec{v}\cdot(\hat{n}_{21}-\hat{n}_{13}))},
\nonumber\\
&\transfer_{\rm E,13}\trls{13}=-2\sin(2\phim)\sin({\phim})e^{i{\phim}(-4-\vec{v}\cdot\hat{n}_{13})}\nonumber\\
&~\times\big[{\rm sinc}\left({\phim}\doppler{13}\right)-{\rm sinc}\left({\phim}\doppleranti{13}\right)
    \big].
\end{align}

The detector tensor for T channel is calculated as
\begin{align}
    \detector{\rm T}=\frac{1}{2\sqrt{3}} \big[&n^i_{21}n^j_{21}\transfer_{\rm T,21}\trls{21}+ n^i_{32}n^j_{32}\transfer_{\rm T,32}\trls{32}\nonumber\\
    &+n^i_{13}n^j_{13}\transfer_{\rm T,13}\trls{13}
    \big]
    \big],
\end{align}
where the transfer functions are
\begin{align}
&\transfer_{\rm T,21}\trls{21}=-
    \sin(2\phim)\sin({\phim})e^{i{\phim}(-4+ \vec{v}\cdot\hat{n}_{21})}\nonumber\\
    &\times
    \big[
    {\rm sinc}\left({\phim}\doppler{21}\right)
    -{\rm sinc}\left({\phim}\doppleranti{21}\right)\big] ,
  \nonumber  \\
&\transfer_{\rm T,32}\trls{32}=-
    \sin(2\phim)\sin({\phim})e^{i{\phim}(-4+ \vec{v}\cdot(\hat{n}_{21}-\hat{n}_{13}))}\nonumber\\
    &\times
    \big[
    {\rm sinc}\left({\phim}\doppler{32}\right)
    -{\rm sinc}\left({\phim}\doppleranti{32}\right)\big] ,
   \nonumber \\
&\transfer_{\rm T,13}\trls{13}=-
\sin(2\phim)\sin(\phim)e^{i\phim(-4- \vec{v}\cdot\hat{n}_{13})}\nonumber\\
&\times
\big[{\rm sinc}\left(\phim\doppler{13}\right) 
-{\rm sinc}\left(\phim\doppleranti{13}\right)\big].
\end{align}
Note that although the T channel is fully symmetric for three arms. Since we choose $h_{ij}(t,\Vec{x}_1)$ as the signal, similarly as in Sec.~\ref{sec_res_ss}, the transfer functions acquire their asymmetric forms. 

After numerical calculation, the response functions for A channel and E channel are the same, which are shown in Fig.~\ref{fig_res_tdi} together with the T channel. From Sec.~\ref{sec_res_ss} and Fig.~\ref{fig_res_tdi}, we can see that $\zeta$ channel and T channel are similar to each other, with the response of GWs being better than of spin-2 ULDM. In fact, these two channels are related by the following simple relation:
\begin{align}
    \response_{\rm T}=3\sin^2(\pi f L)\sin^2(2\pi f L)\response_{\zeta}.
\end{align}
The reason for the weak response of spin-2 ULDM in T and $\zeta$ channel may be attributed to their symmetric features. This effect, combined with the low velocity of spin-2 ULDM, as discussed in Sec.~\ref{sec_res_ss}, results in insensitive performance of the response function. When examining the directional dependence of the response functions, we find that the A and E channels remain largely insensitive to the direction of the ULDM velocity. In contrast, the symmetric T channel exhibits behavior similar to the symmetric Sagnac $\zeta$ channel. However, because the sensitivity of the T channel is significantly weaker than that of the A and E channels, the optimal $\eta$ sensitivity remains effectively unchanged whether the T channel is included or not. Consequently, our results for the constraints presented in Sec. \ref{sec4} are unaffected.

\bibliography{ref.bib}

@article{LIGOScientific:2016aoc,
    author = "Abbott, B. P. and others",
    collaboration = "LIGO Scientific, Virgo",
    title = "{Observation of Gravitational Waves from a Binary Black Hole Merger}",
    eprint = "1602.03837",
    archivePrefix = "arXiv",
    primaryClass = "gr-qc",
    reportNumber = "LIGO-P150914",
    doi = "10.1103/PhysRevLett.116.061102",
    journal = "Phys. Rev. Lett.",
    volume = "116",
    number = "6",
    pages = "061102",
    year = "2016"
}

@article{LIGOScientific:2017vwq,
    author = "Abbott, B. P. and others",
    collaboration = "LIGO Scientific, Virgo",
    title = "{GW170817: Observation of Gravitational Waves from a Binary Neutron Star Inspiral}",
    eprint = "1710.05832",
    archivePrefix = "arXiv",
    primaryClass = "gr-qc",
    reportNumber = "LIGO-P170817",
    doi = "10.1103/PhysRevLett.119.161101",
    journal = "Phys. Rev. Lett.",
    volume = "119",
    number = "16",
    pages = "161101",
    year = "2017"
}

@article{NANOGrav:2023gor,
    author = "Agazie, Gabriella and others",
    collaboration = "NANOGrav",
    title = "{The NANOGrav 15 yr Data Set: Evidence for a Gravitational-wave Background}",
    eprint = "2306.16213",
    archivePrefix = "arXiv",
    primaryClass = "astro-ph.HE",
    doi = "10.3847/2041-8213/acdac6",
    journal = "Astrophys. J. Lett.",
    volume = "951",
    number = "1",
    pages = "L8",
    year = "2023"
}

@article{EPTA:2023fyk,
    author = "Antoniadis, J. and others",
    collaboration = "EPTA, InPTA:",
    title = "{The second data release from the European Pulsar Timing Array - III. Search for gravitational wave signals}",
    eprint = "2306.16214",
    archivePrefix = "arXiv",
    primaryClass = "astro-ph.HE",
    doi = "10.1051/0004-6361/202346844",
    journal = "Astron. Astrophys.",
    volume = "678",
    pages = "A50",
    year = "2023"
}

@article{Reardon:2023gzh,
    author = "Reardon, Daniel J. and others",
    title = "{Search for an Isotropic Gravitational-wave Background with the Parkes Pulsar Timing Array}",
    eprint = "2306.16215",
    archivePrefix = "arXiv",
    primaryClass = "astro-ph.HE",
    doi = "10.3847/2041-8213/acdd02",
    journal = "Astrophys. J. Lett.",
    volume = "951",
    number = "1",
    pages = "L6",
    year = "2023"
}

@article{Xu:2023wog,
    author = "Xu, Heng and others",
    title = "{Searching for the Nano-Hertz Stochastic Gravitational Wave Background with the Chinese Pulsar Timing Array Data Release I}",
    eprint = "2306.16216",
    archivePrefix = "arXiv",
    primaryClass = "astro-ph.HE",
    doi = "10.1088/1674-4527/acdfa5",
    journal = "Res. Astron. Astrophys.",
    volume = "23",
    number = "7",
    pages = "075024",
    year = "2023"
}

@article{BICEP:2021xfz,
    author = "Ade, P. A. R. and others",
    collaboration = "BICEP, Keck",
    title = "{Improved Constraints on Primordial Gravitational Waves using Planck, WMAP, and BICEP/Keck Observations through the 2018 Observing Season}",
    eprint = "2110.00483",
    archivePrefix = "arXiv",
    primaryClass = "astro-ph.CO",
    doi = "10.1103/PhysRevLett.127.151301",
    journal = "Phys. Rev. Lett.",
    volume = "127",
    number = "15",
    pages = "151301",
    year = "2021"
}

@article{Li:2017drr,
    author = "Li, Hong and others",
    title = "{Probing Primordial Gravitational Waves: Ali CMB Polarization Telescope}",
    eprint = "1710.03047",
    archivePrefix = "arXiv",
    primaryClass = "astro-ph.CO",
    doi = "10.1093/nsr/nwy019",
    journal = "Natl. Sci. Rev.",
    volume = "6",
    number = "1",
    pages = "145--154",
    year = "2019"
}

@article{Wang:2024gko,
    author = "Wang, Deng",
    title = "{Primordial Gravitational Waves 2024}",
    eprint = "2407.02714",
    archivePrefix = "arXiv",
    primaryClass = "astro-ph.CO",
    month = "7",
    year = "2024",
    journal = ""
}

@article{Adams:2004pk,
    author = "Adams, A. W. and Bloom, J. S.",
    title = "{Direct detection of dark matter with space-based laser interferometers}",
    eprint = "astro-ph/0405266",
    archivePrefix = "arXiv",
    reportNumber = "HUTP-04-A021",
    month = "5",
    year = "2004",
    journal = ""
}

@article{Hall:2016usm,
    author = {Hall, Evan D. and Adhikari, Rana X. and Frolov, Valery V. and M\"uller, Holger and Pospelov, Maxim and Adhikari, Rana X},
    title = "{Laser Interferometers as Dark Matter Detectors}",
    eprint = "1605.01103",
    archivePrefix = "arXiv",
    primaryClass = "gr-qc",
    doi = "10.1103/PhysRevD.98.083019",
    journal = "Phys. Rev. D",
    volume = "98",
    number = "8",
    pages = "083019",
    year = "2018"
}

@article{Pierce:2018xmy,
    author = "Pierce, Aaron and Riles, Keith and Zhao, Yue",
    title = "{Searching for Dark Photon Dark Matter with Gravitational Wave Detectors}",
    eprint = "1801.10161",
    archivePrefix = "arXiv",
    primaryClass = "hep-ph",
    reportNumber = "LCTP-18-04",
    doi = "10.1103/PhysRevLett.121.061102",
    journal = "Phys. Rev. Lett.",
    volume = "121",
    number = "6",
    pages = "061102",
    year = "2018"
}

@article{Morisaki:2018htj,
    author = "Morisaki, Soichiro and Suyama, Teruaki",
    title = "{Detectability of ultralight scalar field dark matter with gravitational-wave detectors}",
    eprint = "1811.05003",
    archivePrefix = "arXiv",
    primaryClass = "hep-ph",
    reportNumber = "RESCEU-15/18",
    doi = "10.1103/PhysRevD.100.123512",
    journal = "Phys. Rev. D",
    volume = "100",
    number = "12",
    pages = "123512",
    year = "2019"
}

@article{Nagano:2019rbw,
    author = "Nagano, Koji and Fujita, Tomohiro and Michimura, Yuta and Obata, Ippei",
    title = "{Axion Dark Matter Search with Interferometric Gravitational Wave Detectors}",
    eprint = "1903.02017",
    archivePrefix = "arXiv",
    primaryClass = "hep-ph",
    doi = "10.1103/PhysRevLett.123.111301",
    journal = "Phys. Rev. Lett.",
    volume = "123",
    number = "11",
    pages = "111301",
    year = "2019"
}

@article{Guo:2019ker,
    author = "Guo, Huai-Ke and Riles, Keith and Yang, Feng-Wei and Zhao, Yue",
    title = "{Searching for Dark Photon Dark Matter in LIGO O1 Data}",
    eprint = "1905.04316",
    archivePrefix = "arXiv",
    primaryClass = "hep-ph",
    doi = "10.1038/s42005-019-0255-0",
    journal = "Commun. Phys.",
    volume = "2",
    pages = "155",
    year = "2019"
}

@article{Grote:2019uvn,
    author = "Grote, H. and Stadnik, Y. V.",
    title = "{Novel signatures of dark matter in laser-interferometric gravitational-wave detectors}",
    eprint = "1906.06193",
    archivePrefix = "arXiv",
    primaryClass = "astro-ph.IM",
    doi = "10.1103/PhysRevResearch.1.033187",
    journal = "Phys. Rev. Res.",
    volume = "1",
    number = "3",
    pages = "033187",
    year = "2019"
}

@article{Michimura:2020vxn,
    author = "Michimura, Yuta and Fujita, Tomohiro and Morisaki, Soichiro and Nakatsuka, Hiromasa and Obata, Ippei",
    title = "{Ultralight vector dark matter search with auxiliary length channels of gravitational wave detectors}",
    eprint = "2008.02482",
    archivePrefix = "arXiv",
    primaryClass = "hep-ph",
    reportNumber = "JGW-P2011867",
    doi = "10.1103/PhysRevD.102.102001",
    journal = "Phys. Rev. D",
    volume = "102",
    number = "10",
    pages = "102001",
    year = "2020"
}

@article{Miller:2020vsl,
    author = "Miller, Andrew L. and others",
    title = "{Probing new light gauge bosons with gravitational-wave interferometers using an adapted semicoherent method}",
    eprint = "2010.01925",
    archivePrefix = "arXiv",
    primaryClass = "astro-ph.IM",
    doi = "10.1103/PhysRevD.103.103002",
    journal = "Phys. Rev. D",
    volume = "103",
    number = "10",
    pages = "103002",
    year = "2021"
}

@article{Morisaki:2020gui,
    author = "Morisaki, Soichiro and Fujita, Tomohiro and Michimura, Yuta and Nakatsuka, Hiromasa and Obata, Ippei",
    title = "{Improved sensitivity of interferometric gravitational wave detectors to ultralight vector dark matter from the finite light-traveling time}",
    eprint = "2011.03589",
    archivePrefix = "arXiv",
    primaryClass = "hep-ph",
    doi = "10.1103/PhysRevD.103.L051702",
    journal = "Phys. Rev. D",
    volume = "103",
    number = "5",
    pages = "L051702",
    year = "2021"
}

@article{Armaleo:2020efr,
    author = "Armaleo, Juan Manuel and L\'opez Nacir, Diana and Urban, Federico R.",
    title = "{Searching for spin-2 ULDM with gravitational waves interferometers}",
    eprint = "2012.13997",
    archivePrefix = "arXiv",
    primaryClass = "astro-ph.CO",
    doi = "10.1088/1475-7516/2021/04/053",
    journal = "JCAP",
    volume = "04",
    pages = "053",
    year = "2021"
}

@article{Vermeulen:2021epa,
    author = "Vermeulen, Sander M. and others",
    title = "{Direct limits for scalar field dark matter from a gravitational-wave detector}",
    eprint = "2103.03783",
    archivePrefix = "arXiv",
    primaryClass = "gr-qc",
    doi = "10.1038/s41586-021-04031-y",
    month = "3",
    year = "2021",
    journal = ""
}

@article{LIGOScientific:2021ffg,
    author = "Abbott, R. and others",
    collaboration = "LIGO Scientific, KAGRA, Virgo",
    title = "{Constraints on dark photon dark matter using data from LIGO\textquoteright{}s and Virgo\textquoteright{}s third observing run}",
    eprint = "2105.13085",
    archivePrefix = "arXiv",
    primaryClass = "astro-ph.CO",
    reportNumber = "LIGO-P2100098",
    doi = "10.1103/PhysRevD.105.063030",
    journal = "Phys. Rev. D",
    volume = "105",
    number = "6",
    pages = "063030",
    year = "2022",
    note = "[Erratum: Phys.Rev.D 109, 089902 (2024)]"
}

@article{Nagano:2021kwx,
    author = "Nagano, Koji and Nakatsuka, Hiromasa and Morisaki, Soichiro and Fujita, Tomohiro and Michimura, Yuta and Obata, Ippei",
    title = "{Axion dark matter search using arm cavity transmitted beams of gravitational wave detectors}",
    eprint = "2106.06800",
    archivePrefix = "arXiv",
    primaryClass = "hep-ph",
    doi = "10.1103/PhysRevD.104.062008",
    journal = "Phys. Rev. D",
    volume = "104",
    number = "6",
    pages = "062008",
    year = "2021"
}

@article{Aiello:2021wlp,
    author = "Aiello, Lorenzo and Richardson, Jonathan W. and Vermeulen, Sander M. and Grote, Hartmut and Hogan, Craig and Kwon, Ohkyung and Stoughton, Chris",
    title = "{Constraints on Scalar Field Dark Matter from Colocated Michelson Interferometers}",
    eprint = "2108.04746",
    archivePrefix = "arXiv",
    primaryClass = "gr-qc",
    reportNumber = "FERMILAB-PUB-21-352-SCD-T",
    doi = "10.1103/PhysRevLett.128.121101",
    journal = "Phys. Rev. Lett.",
    volume = "128",
    number = "12",
    pages = "121101",
    year = "2022"
}

@article{Chen:2021apc,
    author = "Chen, Chuan-Ren and Nugroho, Chrisna Setyo",
    title = "{Detection prospects of dark matter in the Einstein Telescope}",
    eprint = "2111.11014",
    archivePrefix = "arXiv",
    primaryClass = "hep-ph",
    doi = "10.1103/PhysRevD.105.083001",
    journal = "Phys. Rev. D",
    volume = "105",
    number = "8",
    pages = "083001",
    year = "2022"
}

@article{Miller:2022wxu,
    author = "Miller, Andrew L. and Badaracco, Francesca and Palomba, Cristiano",
    collaboration = "LIGO Scientific, Virgo, KAGRA",
    title = "{Distinguishing between dark-matter interactions with gravitational-wave detectors}",
    eprint = "2204.03814",
    archivePrefix = "arXiv",
    primaryClass = "astro-ph.IM",
    doi = "10.1103/PhysRevD.105.103035",
    journal = "Phys. Rev. D",
    volume = "105",
    number = "10",
    pages = "103035",
    year = "2022"
}

@article{Hall:2022zvi,
    author = "Hall, Evan and Aggarwal, Nancy",
    title = "{Advanced LIGO, LISA, and Cosmic Explorer as dark matter transducers}",
    eprint = "2210.17487",
    archivePrefix = "arXiv",
    primaryClass = "hep-ex",
    month = "10",
    year = "2022",
    journal = ""
}

@article{Ismail:2022ukp,
    author = "Ismail, M. Afif and Nugroho, Chrisna Setyo and Wong, Henry Tsz-King",
    title = "{Exploring dark photons via a subfrequency laser search in gravitational wave detectors}",
    eprint = "2211.13384",
    archivePrefix = "arXiv",
    primaryClass = "hep-ph",
    doi = "10.1103/PhysRevD.107.082002",
    journal = "Phys. Rev. D",
    volume = "107",
    number = "8",
    pages = "082002",
    year = "2023"
}

@article{Miller:2023kkd,
    author = "Miller, Andrew L. and Mendes, Luis",
    title = "{First search for ultralight dark matter with a space-based gravitational-wave antenna: LISA Pathfinder}",
    eprint = "2301.08736",
    archivePrefix = "arXiv",
    primaryClass = "gr-qc",
    doi = "10.1103/PhysRevD.107.063015",
    journal = "Phys. Rev. D",
    volume = "107",
    number = "6",
    pages = "063015",
    year = "2023"
}

@article{Fukusumi:2023kqd,
    author = "Fukusumi, Koki and Morisaki, Soichiro and Suyama, Teruaki",
    title = "{Upper limit on scalar field dark matter from LIGO-Virgo third observation run}",
    eprint = "2303.13088",
    archivePrefix = "arXiv",
    primaryClass = "hep-ph",
    doi = "10.1103/PhysRevD.108.095054",
    journal = "Phys. Rev. D",
    volume = "108",
    number = "9",
    pages = "095054",
    year = "2023"
}

@article{Kim:2023pkx,
    author = "Kim, Hyungjin",
    title = "{Gravitational interaction of ultralight dark matter with interferometers}",
    eprint = "2306.13348",
    archivePrefix = "arXiv",
    primaryClass = "hep-ph",
    reportNumber = "DESY-23-085",
    doi = "10.1088/1475-7516/2023/12/018",
    journal = "JCAP",
    volume = "12",
    pages = "018",
    year = "2023"
}

@article{Yu:2023iog,
    author = "Yu, Jiang-Chuan and Yao, Yue-Hui and Tang, Yong and Wu, Yue-Liang",
    title = "{Sensitivity of space-based gravitational-wave interferometers to ultralight bosonic fields and dark matter}",
    eprint = "2307.09197",
    archivePrefix = "arXiv",
    primaryClass = "gr-qc",
    doi = "10.1103/PhysRevD.108.083007",
    journal = "Phys. Rev. D",
    volume = "108",
    number = "8",
    pages = "083007",
    year = "2023"
}

@article{Heisenberg:2023urf,
    author = "Heisenberg, Lavinia and Maibach, David and Veske, Do\u{g}a",
    title = "{Searching for topological dark matter in LIGO data}",
    eprint = "2309.05093",
    archivePrefix = "arXiv",
    primaryClass = "gr-qc",
    doi = "10.1103/PhysRevD.110.055037",
    journal = "Phys. Rev. D",
    volume = "110",
    number = "5",
    pages = "055037",
    year = "2024"
}

@article{Frerick:2023xnf,
    author = "Frerick, Jonas and Jaeckel, Joerg and Kahlhoefer, Felix and Schmidt-Hoberg, Kai",
    title = "{Riding the dark matter wave: Novel limits on general dark photons from LISA Pathfinder}",
    eprint = "2310.06017",
    archivePrefix = "arXiv",
    primaryClass = "hep-ph",
    reportNumber = "TTP23-044, DESY-23-149",
    doi = "10.1016/j.physletb.2023.138328",
    journal = "Phys. Lett. B",
    volume = "848",
    pages = "138328",
    year = "2024"
}

@article{Manita:2023mnc,
    author = "Manita, Yusuke and Takeda, Hiroki and Aoki, Katsuki and Fujita, Tomohiro and Mukohyama, Shinji",
    title = "{Exploring the spin of ultralight dark matter with gravitational wave detectors}",
    eprint = "2310.10646",
    archivePrefix = "arXiv",
    primaryClass = "hep-ph",
    reportNumber = "KUNS-2979, YITP-23-122, IPMU23-0034",
    doi = "10.1103/PhysRevD.109.095012",
    journal = "Phys. Rev. D",
    volume = "109",
    number = "9",
    pages = "095012",
    year = "2024"
}

@article{Gottel:2024cfj,
    author = {G\"ottel, Alexandre S. and Ejlli, Aldo and Karan, Kanioar and Vermeulen, Sander M. and Aiello, Lorenzo and Raymond, Vivien and Grote, Hartmut},
    title = "{Searching for Scalar Field Dark Matter with LIGO}",
    eprint = "2401.18076",
    archivePrefix = "arXiv",
    primaryClass = "astro-ph.CO",
    doi = "10.1103/PhysRevLett.133.101001",
    journal = "Phys. Rev. Lett.",
    volume = "133",
    number = "10",
    pages = "101001",
    year = "2024"
}

@article{KAGRA:2024ipf,
    author = "Abac, A. G. and others",
    collaboration = "KAGRA, LIGO Scientific, VIRGO",
    title = "{Ultralight vector dark matter search using data from the KAGRA O3GK run}",
    eprint = "2403.03004",
    archivePrefix = "arXiv",
    primaryClass = "astro-ph.CO",
    reportNumber = "LIGO-P2300250",
    doi = "10.1103/PhysRevD.110.042001",
    journal = "Phys. Rev. D",
    volume = "110",
    number = "4",
    pages = "042001",
    year = "2024"
}

@article{Nguyen:2024fpq,
    author = "Nguyen, Quynh Lan and Miller, Andrew L.",
    title = "{Dark Matter and its Effect on Gravitational Wave Signal}",
    doi = "10.22323/1.449.0132",
    journal = "PoS",
    volume = "EPS-HEP2023",
    pages = "132",
    year = "2024"
}

@article{Yao:2024fie,
    author = "Yao, Yue-Hui and Tang, Yong",
    title = "{Probing Stochastic Ultralight Dark Matter with Space-based Gravitational-Wave Interferometers}",
    eprint = "2404.01494",
    archivePrefix = "arXiv",
    primaryClass = "hep-ph",
    month = "4",
    year = "2024",
    journal = ""
}

@article{Yu:2024enm,
    author = "Yu, Jiang-Chuan and Cao, Yan and Tang, Yong and Wu, Yue-Liang",
    title = "{Detecting ultralight dark matter gravitationally with laser interferometers in space}",
    eprint = "2404.04333",
    archivePrefix = "arXiv",
    primaryClass = "hep-ph",
    doi = "10.1103/PhysRevD.110.023025",
    journal = "Phys. Rev. D",
    volume = "110",
    number = "2",
    pages = "023025",
    year = "2024"
}

@article{Ferreira:2020fam,
    author = "Ferreira, Elisa G. M.",
    title = "{Ultra-light dark matter}",
    eprint = "2005.03254",
    archivePrefix = "arXiv",
    primaryClass = "astro-ph.CO",
    doi = "10.1007/s00159-021-00135-6",
    journal = "Astron. Astrophys. Rev.",
    volume = "29",
    number = "1",
    pages = "7",
    year = "2021"
}

@article{Hui:2021tkt,
    author = "Hui, Lam",
    title = "{Wave Dark Matter}",
    eprint = "2101.11735",
    archivePrefix = "arXiv",
    primaryClass = "astro-ph.CO",
    doi = "10.1146/annurev-astro-120920-010024",
    journal = "Ann. Rev. Astron. Astrophys.",
    volume = "59",
    pages = "247--289",
    year = "2021"
}

@article{Aghaie:2023lan,
    author = "Aghaie, Mohammad and Armando, Giovanni and Dondarini, Alessandro and Panci, Paolo",
    title = "{Bounds on ultralight dark matter from NANOGrav}",
    eprint = "2308.04590",
    archivePrefix = "arXiv",
    primaryClass = "astro-ph.CO",
    doi = "10.1103/PhysRevD.109.103030",
    journal = "Phys. Rev. D",
    volume = "109",
    number = "10",
    pages = "103030",
    year = "2024"
}

@article{Bromley:2023yfi,
    author = "Bromley, Benjamin C. and Sandick, Pearl and Shams Es Haghi, Barmak",
    title = "{Supermassive black hole binaries in ultralight dark matter}",
    eprint = "2311.18013",
    archivePrefix = "arXiv",
    primaryClass = "astro-ph.GA",
    reportNumber = "UTWI-36-2023",
    doi = "10.1103/PhysRevD.110.023517",
    journal = "Phys. Rev. D",
    volume = "110",
    number = "2",
    pages = "023517",
    year = "2024"
}

@article{Fell:2023mtf,
    author = "Fell, Shaun David Brocus and Heisenberg, Lavinia and Veske, Do\u{g}a",
    title = "{Detecting fundamental vector fields with LISA}",
    eprint = "2304.14129",
    archivePrefix = "arXiv",
    primaryClass = "gr-qc",
    doi = "10.1103/PhysRevD.108.083010",
    journal = "Phys. Rev. D",
    volume = "108",
    number = "8",
    pages = "083010",
    year = "2023"
}

@article{Brito:2015oca,
    author = "Brito, Richard and Cardoso, Vitor and Pani, Paolo",
    title = "{Superradiance}: {New Frontiers in Black Hole
Physics}",
    eprint = "1501.06570",
    archivePrefix = "arXiv",
    primaryClass = "gr-qc",
    doi = "10.1007/978-3-319-19000-6",
    journal = "Lect. Notes Phys.",
    volume = "906",
    pages = "pp.1--237",
    year = "2015"
}

@article{Brito:2017wnc,
    author = "Brito, Richard and Ghosh, Shrobana and Barausse, Enrico and Berti, Emanuele and Cardoso, Vitor and Dvorkin, Irina and Klein, Antoine and Pani, Paolo",
    title = "{Stochastic and resolvable gravitational waves from ultralight bosons}",
    eprint = "1706.05097",
    archivePrefix = "arXiv",
    primaryClass = "gr-qc",
    doi = "10.1103/PhysRevLett.119.131101",
    journal = "Phys. Rev. Lett.",
    volume = "119",
    number = "13",
    pages = "131101",
    year = "2017"
}

@article{Brito:2017zvb,
    author = "Brito, Richard and Ghosh, Shrobana and Barausse, Enrico and Berti, Emanuele and Cardoso, Vitor and Dvorkin, Irina and Klein, Antoine and Pani, Paolo",
    title = "{Gravitational wave searches for ultralight bosons with LIGO and LISA}",
    eprint = "1706.06311",
    archivePrefix = "arXiv",
    primaryClass = "gr-qc",
    doi = "10.1103/PhysRevD.96.064050",
    journal = "Phys. Rev. D",
    volume = "96",
    number = "6",
    pages = "064050",
    year = "2017"
}

@article{Isi:2018pzk,
    author = "Isi, Maximiliano and Sun, Ling and Brito, Richard and Melatos, Andrew",
    title = "{Directed searches for gravitational waves from ultralight bosons}",
    eprint = "1810.03812",
    archivePrefix = "arXiv",
    primaryClass = "gr-qc",
    reportNumber = "LIGO-P1800270",
    doi = "10.1103/PhysRevD.99.084042",
    journal = "Phys. Rev. D",
    volume = "99",
    number = "8",
    pages = "084042",
    year = "2019",
    note = "[Erratum: Phys.Rev.D 102, 049901 (2020)]"
}

@article{Yuan:2021ebu,
    author = "Yuan, Chen and Brito, Richard and Cardoso, Vitor",
    title = "{Probing ultralight dark matter with future ground-based gravitational-wave detectors}",
    eprint = "2106.00021",
    archivePrefix = "arXiv",
    primaryClass = "gr-qc",
    doi = "10.1103/PhysRevD.104.044011",
    journal = "Phys. Rev. D",
    volume = "104",
    number = "4",
    pages = "044011",
    year = "2021"
}

@article{deRham:2010ik,
    author = "de Rham, Claudia and Gabadadze, Gregory",
    title = "{Generalization of the Fierz-Pauli Action}",
    eprint = "1007.0443",
    archivePrefix = "arXiv",
    primaryClass = "hep-th",
    reportNumber = "NYU-TH-06-13-10",
    doi = "10.1103/PhysRevD.82.044020",
    journal = "Phys. Rev. D",
    volume = "82",
    pages = "044020",
    year = "2010"
}

@article{deRham:2010kj,
    author = "de Rham, Claudia and Gabadadze, Gregory and Tolley, Andrew J.",
    title = "{Resummation of Massive Gravity}",
    eprint = "1011.1232",
    archivePrefix = "arXiv",
    primaryClass = "hep-th",
    doi = "10.1103/PhysRevLett.106.231101",
    journal = "Phys. Rev. Lett.",
    volume = "106",
    pages = "231101",
    year = "2011"
}

@article{Yuan:2022bem,
    author = "Yuan, Chen and Jiang, Yang and Huang, Qing-Guo",
    title = "{Constraints on an ultralight scalar boson from Advanced LIGO and Advanced Virgo\textquoteright{}s first three observing runs using the stochastic gravitational-wave background}",
    eprint = "2204.03482",
    archivePrefix = "arXiv",
    primaryClass = "astro-ph.CO",
    doi = "10.1103/PhysRevD.106.023020",
    journal = "Phys. Rev. D",
    volume = "106",
    number = "2",
    pages = "023020",
    year = "2022"
}

@article{Guo:2023gfc,
    author = "Guo, Rong-Zhen and Jiang, Yang and Huang, Qing-Guo",
    title = "{Probing ultralight tensor dark matter with the stochastic gravitational-wave background from advanced LIGO and Virgo's first three observing runs}",
    eprint = "2312.16435",
    archivePrefix = "arXiv",
    primaryClass = "astro-ph.CO",
    doi = "10.1088/1475-7516/2024/04/053",
    journal = "JCAP",
    volume = "04",
    pages = "053",
    year = "2024"
}

@article{Xie:2024xex,
    author = "Xie, Yiqi and Chung, Adrian Ka-Wai and Sotiriou, Thomas P. and Yunes, Nicol\'as",
    title = "{Bayesian search of massive scalar fields from LIGO-Virgo-KAGRA binaries}",
    eprint = "2410.14801",
    archivePrefix = "arXiv",
    primaryClass = "gr-qc",
    month = "10",
    year = "2024",
    journal = ""
}

@article{Armaleo:2020yml,
    author = "Armaleo, Juan Manuel and L\'opez Nacir, Diana and Urban, Federico R.",
    title = "{Pulsar timing array constraints on spin-2 ULDM}",
    eprint = "2005.03731",
    archivePrefix = "arXiv",
    primaryClass = "astro-ph.CO",
    doi = "10.1088/1475-7516/2020/09/031",
    journal = "JCAP",
    volume = "09",
    pages = "031",
    year = "2020"
}

@article{Sun:2021yra,
    author = "Sun, Sichun and Yang, Xing-Yu and Zhang, Yun-Long",
    title = "{Pulsar timing residual induced by wideband ultralight dark matter with spin 0,1,2}",
    eprint = "2112.15593",
    archivePrefix = "arXiv",
    primaryClass = "astro-ph.CO",
    doi = "10.1103/PhysRevD.106.066006",
    journal = "Phys. Rev. D",
    volume = "106",
    number = "6",
    pages = "066006",
    year = "2022"
}

@article{Unal:2022ooa,
    author = "Unal, Caner and Urban, Federico R. and Kovetz, Ely D.",
    title = "{Probing ultralight scalar, vector and tensor dark matter with pulsar timing arrays}",
    eprint = "2209.02741",
    archivePrefix = "arXiv",
    primaryClass = "astro-ph.CO",
    doi = "10.1016/j.physletb.2024.138830",
    journal = "Phys. Lett. B",
    volume = "855",
    pages = "138830",
    year = "2024"
}

@article{Wu:2023dnp,
    author = "Wu, Yu-Mei and Chen, Zu-Cheng and Huang, Qing-Guo",
    title = "{Pulsar timing residual induced by ultralight tensor dark matter}",
    eprint = "2305.08091",
    archivePrefix = "arXiv",
    primaryClass = "hep-ph",
    doi = "10.1088/1475-7516/2023/09/021",
    journal = "JCAP",
    volume = "09",
    pages = "021",
    year = "2023"
}

@article{LISA:2017pwj,
    author = "Amaro-Seoane, Pau and others",
    collaboration = "LISA",
    title = "{Laser Interferometer Space Antenna}",
    eprint = "1702.00786",
    archivePrefix = "arXiv",
    primaryClass = "astro-ph.IM",
    month = "2",
    year = "2017",
    journal = ""
}

@article{Hu:2017mde,
    author = "Hu, Wen-Rui and Wu, Yue-Liang",
    title = "{The Taiji Program in Space for gravitational wave physics and the nature of gravity}",
    doi = "10.1093/nsr/nwx116",
    journal = "Natl. Sci. Rev.",
    volume = "4",
    number = "5",
    pages = "685--686",
    year = "2017"
}

@article{TianQin:2015yph,
    author = "Luo, Jun and others",
    collaboration = "TianQin",
    title = "{TianQin: a space-borne gravitational wave detector}",
    eprint = "1512.02076",
    archivePrefix = "arXiv",
    primaryClass = "astro-ph.IM",
    doi = "10.1088/0264-9381/33/3/035010",
    journal = "Class. Quant. Grav.",
    volume = "33",
    number = "3",
    pages = "035010",
    year = "2016"
}

@article{Tinto:2004wu,
    author = "Tinto, Massimo and Dhurandhar, Sanjeev V.",
    title = "{Time-Delay Interferometry}",
    eprint = "gr-qc/0409034",
    archivePrefix = "arXiv",
    doi = "10.12942/lrr-2005-4",
    journal = "Living Rev. Rel.",
    volume = "8",
    pages = "4",
    year = "2005"
}

@article{Dubovsky:2004ud,
    author = "Dubovsky, S. L. and Tinyakov, P. G. and Tkachev, I. I.",
    title = "{Massive graviton as a testable cold dark matter candidate}",
    eprint = "hep-th/0411158",
    archivePrefix = "arXiv",
    doi = "10.1103/PhysRevLett.94.181102",
    journal = "Phys. Rev. Lett.",
    volume = "94",
    pages = "181102",
    year = "2005"
}

@article{Yamashita:2014fga,
    author = "Yamashita, Yasuho and De Felice, Antonio and Tanaka, Takahiro",
    title = "{Appearance of Boulware\textendash{}Deser ghost in bigravity with doubly coupled matter}",
    eprint = "1408.0487",
    archivePrefix = "arXiv",
    primaryClass = "hep-th",
    reportNumber = "KUNS-2508, YITP-15-37",
    doi = "10.1142/S0218271814430032",
    journal = "Int. J. Mod. Phys. D",
    volume = "23",
    pages = "1443003",
    year = "2014"
}

@article{Hassan:2011zd,
    author = "Hassan, S. F. and Rosen, Rachel A.",
    title = "{Bimetric Gravity from Ghost-free Massive Gravity}",
    eprint = "1109.3515",
    archivePrefix = "arXiv",
    primaryClass = "hep-th",
    doi = "10.1007/JHEP02(2012)126",
    journal = "JHEP",
    volume = "02",
    pages = "126",
    year = "2012"
}

@article{Aoki:2016zgp,
    author = "Aoki, Katsuki and Mukohyama, Shinji",
    title = "{Massive gravitons as dark matter and gravitational waves}",
    eprint = "1604.06704",
    archivePrefix = "arXiv",
    primaryClass = "hep-th",
    doi = "10.1103/PhysRevD.94.024001",
    journal = "Phys. Rev. D",
    volume = "94",
    number = "2",
    pages = "024001",
    year = "2016"
}

@article{Babichev:2016hir,
    author = {Babichev, Eugeny and Marzola, Luca and Raidal, Martti and Schmidt-May, Angnis and Urban, Federico and Veerm\"ae, Hardi and von Strauss, Mikael},
    title = "{Bigravitational origin of dark matter}",
    eprint = "1604.08564",
    archivePrefix = "arXiv",
    primaryClass = "hep-ph",
    reportNumber = "LPT-ORSAY-16-75",
    doi = "10.1103/PhysRevD.94.084055",
    journal = "Phys. Rev. D",
    volume = "94",
    number = "8",
    pages = "084055",
    year = "2016"
}

@article{Babichev:2016bxi,
    author = {Babichev, Eugeny and Marzola, Luca and Raidal, Martti and Schmidt-May, Angnis and Urban, Federico and Veerm\"ae, Hardi and von Strauss, Mikael},
    title = "{Heavy spin-2 Dark Matter}",
    eprint = "1607.03497",
    archivePrefix = "arXiv",
    primaryClass = "hep-th",
    reportNumber = "LPT-ORSAY-16-60",
    doi = "10.1088/1475-7516/2016/09/016",
    journal = "JCAP",
    volume = "09",
    pages = "016",
    year = "2016"
}

@article{Aoki:2017cnz,
    author = "Aoki, Katsuki and Maeda, Kei-ichi",
    title = "{Condensate of Massive Graviton and Dark Matter}",
    eprint = "1707.05003",
    archivePrefix = "arXiv",
    primaryClass = "hep-th",
    reportNumber = "WU-AP-1702-17",
    doi = "10.1103/PhysRevD.97.044002",
    journal = "Phys. Rev. D",
    volume = "97",
    number = "4",
    pages = "044002",
    year = "2018"
}

@article{Marzola:2017lbt,
    author = "Marzola, Luca and Raidal, Martti and Urban, Federico R.",
    title = "{Oscillating Spin-2 Dark Matter}",
    eprint = "1708.04253",
    archivePrefix = "arXiv",
    primaryClass = "hep-ph",
    doi = "10.1103/PhysRevD.97.024010",
    journal = "Phys. Rev. D",
    volume = "97",
    number = "2",
    pages = "024010",
    year = "2018"
}

@article{Xia:2023hov,
    author = "Xia, Zi-Qing and Tang, Tian-Peng and Huang, Xiaoyuan and Yuan, Qiang and Fan, Yi-Zhong",
    title = "{Constraining ultralight dark matter using the Fermi-LAT pulsar timing array}",
    eprint = "2303.17545",
    archivePrefix = "arXiv",
    primaryClass = "astro-ph.HE",
    doi = "10.1103/PhysRevD.107.L121302",
    journal = "Phys. Rev. D",
    volume = "107",
    number = "12",
    pages = "L121302",
    year = "2023"
}

@article{Cai:2024thd,
    author = "Cai, Rong-Gen and Zhang, Jing-Rui and Zhang, Yun-Long",
    title = "{Angular correlation and deformed Hellings-Downs curve from spin-2 ultralight dark matter}",
    eprint = "2402.03984",
    archivePrefix = "arXiv",
    primaryClass = "gr-qc",
    doi = "10.1103/PhysRevD.110.044052",
    journal = "Phys. Rev. D",
    volume = "110",
    number = "4",
    pages = "044052",
    year = "2024"
}

@article{Cornish:2001bb,
    author = "Cornish, Neil J.",
    title = "{Detecting a stochastic gravitational wave background with the Laser Interferometer Space Antenna}",
    eprint = "gr-qc/0106058",
    archivePrefix = "arXiv",
    doi = "10.1103/PhysRevD.65.022004",
    journal = "Phys. Rev. D",
    volume = "65",
    pages = "022004",
    year = "2002"
}

@article{Robson:2018ifk,
    author = "Robson, Travis and Cornish, Neil J. and Liu, Chang",
    title = "{The construction and use of LISA sensitivity curves}",
    eprint = "1803.01944",
    archivePrefix = "arXiv",
    primaryClass = "astro-ph.HE",
    doi = "10.1088/1361-6382/ab1101",
    journal = "Class. Quant. Grav.",
    volume = "36",
    number = "10",
    pages = "105011",
    year = "2019"
}

@article{Fierz:1939ix,
    author = "Fierz, M. and Pauli, W.",
    title = "{On relativistic wave equations for particles of arbitrary spin in an electromagnetic field}",
    doi = "10.1098/rspa.1939.0140",
    journal = "Proc. Roy. Soc. Lond. A",
    volume = "173",
    pages = "211--232",
    year = "1939"
}

@article{Lee:2024oxo,
    author = "Lee, Vincent S. H. and Zurek, Kathryn M.",
    title = "{Proper Time Observables of General Gravitational Perturbations in Laser Interferometry-based Gravitational Wave Detectors}",
    eprint = "2408.03363",
    archivePrefix = "arXiv",
    primaryClass = "hep-ph",
    reportNumber = "CALT-TH-2024-025",
    month = "8",
    year = "2024",
    journal = ""
}

@article{Tsai:2023zza,
    author = "Tsai, Yu-Dai and Farnocchia, Davide and Micheli, Marco and Vagnozzi, Sunny and Visinelli, Luca",
    title = "{Constraints on fifth forces and ultralight dark matter from OSIRIS-REx target asteroid Bennu}",
    eprint = "2309.13106",
    archivePrefix = "arXiv",
    primaryClass = "hep-ph",
    reportNumber = "UCI-HEP-TR-2023-04, FERMILAB-PUB-23-538-T-V",
    doi = "10.1038/s42005-024-01779-3",
    journal = "Commun. Phys.",
    volume = "7",
    number = "1",
    pages = "311",
    year = "2024"
}

@article{Hohmann:2017uxe,
    author = "Hohmann, Manuel",
    title = "{Post-Newtonian parameter \ensuremath{\gamma} and the deflection of light in ghost-free massive bimetric gravity}",
    eprint = "1701.07700",
    archivePrefix = "arXiv",
    primaryClass = "gr-qc",
    doi = "10.1103/PhysRevD.95.124049",
    journal = "Phys. Rev. D",
    volume = "95",
    number = "12",
    pages = "124049",
    year = "2017"
}

@article{Sereno:2006mw,
    author = "Sereno, Mauro and Jetzer, Ph.",
    title = "{Dark matter vs. modifications of the gravitational inverse-square law. Results from planetary motion in the solar system}",
    eprint = "astro-ph/0606197",
    archivePrefix = "arXiv",
    doi = "10.1111/j.1365-2966.2006.10670.x",
    journal = "Mon. Not. Roy. Astron. Soc.",
    volume = "371",
    pages = "626--632",
    year = "2006"
}

@book{Maggiore:2018sht,
    author = "Maggiore, Michele",
    title = "Gravitational Waves. Vol. 2: Astrophysics and Cosmology",
    isbn = "978-0-19-857089-9",
    publisher = "Oxford University Press",
    month = "3",
    year = "2018"
}

@article{Blas:2024jyh,
    author = "Blas, Diego and Carlton, John and McCabe, Christopher",
    title = "{Massive graviton dark matter searches with long-baseline atom interferometers}",
    eprint = "2412.14282",
    archivePrefix = "arXiv",
    primaryClass = "hep-ph",
    reportNumber = "KCL-PH-TH/2024-69, AION-REPORT/2024-08",
    month = "12",
    year = "2024",
    journal = ""
}

@article{Prince:2002hp,
    author = "Prince, Thomas A. and Tinto, Massimo and Larson, Shane L. and Armstrong, J. W.",
    title = "{The LISA optimal sensitivity}",
    eprint = "gr-qc/0209039",
    archivePrefix = "arXiv",
    doi = "10.1103/PhysRevD.66.122002",
    journal = "Phys. Rev. D",
    volume = "66",
    pages = "122002",
    year = "2002"
}

@article{Mitra:2023sny,
    author = "Mitra, Soumodeep and Chakraborty, Sumanta and Vicente, Rodrigo and Feng, Justin C.",
    title = "{Probing the quantum nature of black holes with ultralight boson environments}",
    eprint = "2312.06783",
    archivePrefix = "arXiv",
    primaryClass = "gr-qc",
    doi = "10.1103/PhysRevD.110.084012",
    journal = "Phys. Rev. D",
    volume = "110",
    number = "8",
    pages = "084012",
    year = "2024"
}

@article{Kolb:2023dzp,
    author = "Kolb, Edward W. and Ling, Siyang and Long, Andrew J. and Rosen, Rachel A.",
    title = "{Cosmological gravitational particle production of massive spin-2 particles}",
    eprint = "2302.04390",
    archivePrefix = "arXiv",
    primaryClass = "astro-ph.CO",
    doi = "10.1007/JHEP05(2023)181",
    journal = "JHEP",
    volume = "05",
    pages = "181",
    year = "2023"
}

@article{Jain:2021pnk,
    author = "Jain, Mudit and Amin, Mustafa A.",
    title = "{Polarized solitons in higher-spin wave dark matter}",
    eprint = "2109.04892",
    archivePrefix = "arXiv",
    primaryClass = "hep-th",
    doi = "10.1103/PhysRevD.105.056019",
    journal = "Phys. Rev. D",
    volume = "105",
    number = "5",
    pages = "056019",
    year = "2022"
}

@article{Chowdhury:2023xvy,
    author = "Chowdhury, Debtosh and Hait, Arpan and Mohanty, Subhendra and Prakash, Suraj",
    title = "{Ultralight dark matter explanation of NANOGrav observations}",
    eprint = "2311.10148",
    archivePrefix = "arXiv",
    primaryClass = "hep-ph",
    doi = "10.1103/PhysRevD.110.083023",
    journal = "Phys. Rev. D",
    volume = "110",
    number = "8",
    pages = "083023",
    year = "2024"
}

@article{Brito:2013wya,
    author = "Brito, Richard and Cardoso, Vitor and Pani, Paolo",
    title = "{Massive spin-2 fields on black hole spacetimes: Instability of the Schwarzschild and Kerr solutions and bounds on the graviton mass}",
    eprint = "1304.6725",
    archivePrefix = "arXiv",
    primaryClass = "gr-qc",
    doi = "10.1103/PhysRevD.88.023514",
    journal = "Phys. Rev. D",
    volume = "88",
    number = "2",
    pages = "023514",
    year = "2013"
}

@article{Brito:2020lup,
    author = "Brito, Richard and Grillo, Sara and Pani, Paolo",
    title = "{Black Hole Superradiant Instability from Ultralight Spin-2 Fields}",
    eprint = "2002.04055",
    archivePrefix = "arXiv",
    primaryClass = "gr-qc",
    doi = "10.1103/PhysRevLett.124.211101",
    journal = "Phys. Rev. Lett.",
    volume = "124",
    number = "21",
    pages = "211101",
    year = "2020"
}

@article{Dias:2023ynv,
    author = "Dias, Oscar J. C. and Lingetti, Giuseppe and Pani, Paolo and Santos, Jorge E.",
    title = "{Black hole superradiant instability for massive spin-2 fields}",
    eprint = "2304.01265",
    archivePrefix = "arXiv",
    primaryClass = "gr-qc",
    doi = "10.1103/PhysRevD.108.L041502",
    journal = "Phys. Rev. D",
    volume = "108",
    number = "4",
    pages = "L041502",
    year = "2023"
}

@article{Manita:2022tkl,
    author = "Manita, Yusuke and Aoki, Katsuki and Fujita, Tomohiro and Mukohyama, Shinji",
    title = "{Spin-2 dark matter from an anisotropic universe in bigravity}",
    eprint = "2211.15873",
    archivePrefix = "arXiv",
    primaryClass = "gr-qc",
    reportNumber = "KUNS-2946, YITP-22-150, IPMU22-0063",
    doi = "10.1103/PhysRevD.107.104007",
    journal = "Phys. Rev. D",
    volume = "107",
    number = "10",
    pages = "104007",
    year = "2023"
}

@article{Aoki:2019snr,
    author = "Aoki, Katsuki and Mukohyama, Shinji",
    title = "{Ghostfree quadratic curvature theories with massive spin-2 and spin-0 particles}",
    eprint = "1907.09690",
    archivePrefix = "arXiv",
    primaryClass = "hep-th",
    reportNumber = "YITP-19-67, IPMU19-0098",
    doi = "10.1103/PhysRevD.100.064061",
    journal = "Phys. Rev. D",
    volume = "100",
    number = "6",
    pages = "064061",
    year = "2019"
}

@article{Arjona:2024cex,
    author = "Arjona, Rub\'en and Nesseris, Savvas and Kuroyanagi, Sachiko",
    title = "{Comparative analysis of the NANOgrav Hellings-Downs as a window into new physics}",
    eprint = "2412.12975",
    archivePrefix = "arXiv",
    primaryClass = "astro-ph.CO",
    reportNumber = "IFT-UAM/CSIC-24-173",
    month = "12",
    year = "2024",
    journal = ""
}

@article{Armaleo:2019gil,
    author = "Armaleo, Juan Manuel and L\'opez Nacir, Diana and Urban, Federico R.",
    title = "{Binary pulsars as probes for spin-2 ultralight dark matter}",
    eprint = "1909.13814",
    archivePrefix = "arXiv",
    primaryClass = "astro-ph.HE",
    doi = "10.1088/1475-7516/2020/01/053",
    journal = "JCAP",
    volume = "01",
    pages = "053",
    year = "2020"
}

@article{Roberts:2018xqn,
    author = "Roberts, B. M. and Blewitt, G. and Dailey, C. and Derevianko, A.",
    title = "{Search for transient ultralight dark matter signatures with networks of precision measurement devices using a Bayesian statistics method}",
    eprint = "1803.10264",
    archivePrefix = "arXiv",
    primaryClass = "astro-ph.IM",
    doi = "10.1103/PhysRevD.97.083009",
    journal = "Phys. Rev. D",
    volume = "97",
    number = "8",
    pages = "083009",
    year = "2018"
}

 \end{document}